\documentclass[aps,prb,floatfix]{revtex4}
\usepackage{epsfig}
\begin{document}
\tighten


\title{Theory of spin-polarized bipolar transport in magnetic {\it p-n} junctions.}

\author{Jaroslav Fabian}
\affiliation{Institute for Theoretical Physics, Karl-Franzens University,
Universit\"atsplatz 5, 8010 Graz, Austria }

\author{Igor \v{Z}uti\'{c} and S. Das Sarma}
\affiliation{Condensed Matter Theory Center,
Department of Physics, University of Maryland at College
Park, College Park, Maryland 20742-4111, USA}


\begin{abstract}
{The interplay between spin and charge transport in electrically and magnetically 
inhomogeneous semiconductor systems is investigated theoretically. In particular, the
theory of spin-polarized bipolar transport in magnetic {\it p-n} 
junctions is formulated, generalizing the classic Shockley model. 
The theory assumes that in the depletion layer the nonequilibrium chemical 
potentials of spin up and spin down carriers are constant
and carrier recombination and spin relaxation are inhibited. 
Under the general conditions of an applied bias and externally injected (source) spin, 
the model formulates analytically carrier and spin transport in 
magnetic {\it p-n} junctions at low bias. The evaluation of the carrier
and spin densities at the depletion layer establishes the necessary boundary conditions for solving
the diffusive transport equations in the bulk regions separately, thus greatly
simplifying the  problem. The carrier and spin density and current
profiles in the bulk regions are calculated and the I-V characteristics 
of the junction are obtained. It is demonstrated that spin injection through 
the depletion layer of a magnetic {\it p-n} junction is not possible unless 
 nonequilibrium spin accumulates 
in the bulk regions--either by external spin injection or by the application of a large bias.
Implications of the theory for majority spin injection across the depletion layer, minority spin
pumping and spin amplification, giant magnetoresistance, spin-voltaic effect, 
biasing electrode spin injection, and magnetic drift in the bulk regions  
are discussed in details, and illustrated using the example of a GaAs based magnetic {\it p-n} junction.
}
\end{abstract}

\maketitle

\section{introduction}
Active control of spin in semiconductors\cite{dassarma01} 
is projected to lead to significant
technological advances, most importantly 
in digital information storage and processing, magnetic recording and sensing, 
and quantum computing.~\cite{loss98,hu00} 
Using semiconductors for spintronic applications--where
spin, in addition to charge, is manipulated to influence electronic 
properties--has several advantages. First, integration of spintronics with 
traditional semiconductor technology calls for employing semiconductors 
(rather than metals) as media for spin control. Second, semiconductors
are versatile materials, not only for their electrical properties,
but also for their spin/magnetic characteristics. Doping control
of electrical and magnetic properties, optical spin
orientation and detection, bipolar (electron and hole) transport, and 
interface properties (charge and spin accumulation and depletion)
leading to device concepts from {\it p-n} junction diodes to 
field-effect transistors, are among the great advantages of 
semiconductors over other candidates for spintronic materials.
By allowing for the active control and manipulation of carrier spin and charge
by electric and magnetic fields as well as by light, semiconductor
spintronics creates the potential for an integrated 
magneto-optoelectronics technology.

A generic semiconductor spintronics scheme involves three steps:
injection of nonequilibrium spin into a 
semiconductor, spin storage, manipulation, and transfer, 
and spin detection. 
Spin injection was historically first accomplished optically, 
by illuminating a semiconductor with circularly polarized 
light--the so called spin orientation.\cite{optical84}
Electrical spin injection (that is, spin injection from a magnetic
electrode, often called simply spin injection)
into semiconductors, while predicted theoretically already in the 
70s,\cite{aronov76} has been demonstrated 
only recently,\cite{oestreich99} 
and realized as an injection from a magnetic 
semiconductor,\cite{fiederling99,ohno99,jonker00}
a ferromagnetic metal,\cite{hammar99,zhu01,dediu02}  
and a ferromagnetic metal/tunnel barrier
contact.\cite{nitta01,hammar01,hanbicki02,motsnyi02,parkin02} 

Once injected, 
nonequilibrium spin survives for a  reasonably
long time when compared to typical relaxation times of momentum and energy
of the injected carriers. Room temperature spin 
relaxation times in semiconductors are typically 
nanoseconds\cite{optical84,fabian99} (compared to sub picosecond time scales 
for momentum and energy relaxation).
Similar in magnitude are only carrier (electron and hole)
recombination times, which are usually between micro- to
nanoseconds. If not in the ballistic regime, transport of spin in a semiconductor can 
be characterized as carrier recombination and spin relaxation limited
drift and diffusion. Spin typically diffuses over micron distances 
from the point of injection, sufficient  
for microelectronics applications.
 Application of large electric fields can
further drag the injected spin over several microns at low temperatures,
as in intrinsic GaAs,\cite{hagele98} and even up to 100 $\mu$m in $n$-doped 
GaAs.\cite{kikkawa99} (As far as the spin diffusion length is concerned, metals
have an advantage: because of the large Fermi velocity, spin
diffusion lengths in metals can be as large as centimeters.) 
Important for device applications are 
studies of spin transport in inhomogeneous semiconductors. It has
already been shown, for example, that  spin phase can be preserved in 
transport across heterostructure interfaces,\cite{malajovich01}
that electron spin can be controlled by bias in semimagnetic 
resonant tunneling diodes,\cite{gruber01} 
and that spin can tunnel through the transition region of tunnel 
diodes.\cite{esaki,zener}
The final step of a generic spintronics scheme
is spin detection. Traditionally, spin in semiconductors has been 
detected optically by observing circular polarization of the recombination light. 
\cite{optical84} Efforts to electrically
detect nonequilibrium spin in semiconductors rely on spin-charge coupling, realized either as 
spin-dependent Schottky barrier transport\cite{bland01,isakovic01}  or as
magnetoresistance\cite{schmidt01} and galvano-voltaic\cite{ganichev01}
effects. 

After the discovery of ferromagnetism in III-V semiconductor 
compounds,\cite{munekata89,ohno92}
the great push for semiconductor spintronics came with the fabrication
of (Ga,Mn)As which is ferromagnetic above 100 K.\cite{ohno98,esch97}
Ferromagnetic semiconductors can not only serve to inject and detect spin 
in all-semiconductor spintronic devices, but can also form
a basis for nonvolatile memory, opening
prospects of integrated, single-chip memory and logic applications
(feasibility of such prospects has been demonstrated by controlling
semiconductor ferromagnetism optically\cite{krenn89,koshihara97} and 
electrically\cite{ohno00}). There is a steady increase in the number of available
ferromagnetic semiconductors, including a first group-IV 
compound GeMn,\cite{park02}
(In,Ga,Mn)As,\cite{slupinski02}  reported room temperature ferromagnets
Mn-doped CdGeP,\cite{CdMnGeP}  GaN,\cite{reed02} and  GaP,\cite{hebard02}
and Co-doped TiO$_2$.\cite{matsumoto01}
    
Closely following the experimental progress, major theoretical efforts                  
have been dedicated to understanding electrical spin injection into 
semiconductors\cite{schmidt00,rashba00,smith01,fert01,hu01,silsbee01,rashba02}
and investigating fundamental issues of spin-polarized transport in semiconductors.
\cite{optical84,dassarma00,tang00,flatte00,zutic01a,zutic01b,zutic02,yu02,martin02,zutic02b}
Another direction for fundamental spintronics theory has been predicting and analyzing various 
spintronics  device architectures for possible technological\cite{krenn89}
applications. The common goal of these studies is devising  
spin valves and structures (typically including one or several magnetic
layers) with maximized magnetoresistance. To this end various spin field-effect 
transistors have been proposed,\cite{datta90,gardelis99,schapers01} 
where the source and drain are ferromagnetic 
electrodes serving to inject and detect spin which is transported
in a (typically) nonmagnetic channel. Spin and charge transport in the
channel are controlled by gate bias through the Rashba 
effect.\cite{rashba60,rashba84} 
Other proposed spintronics device schemes include 
heterostructure spin 
filters\cite{egues98,silva99,kircenzow01,guo01,kiselev01,efros01,hu02,gover01,wrobel01,ramaglia02} 
and spin polarization detectors,\cite{zutic99a} resonant tunneling 
diodes,\cite{petukhov98} 
unipolar magnetic diodes,\cite{flatte01}
quantum-interference mesoscopic schemes\cite{frustaglia01,koshi01,nikolic01}
and various spin emf sources.
\cite{ganichev01,dyakonov71a,dyakonov71b,hirsch99,bhat00,brataas02}

We have recently proposed two spintronics device schemes that take
advantage of {\it bipolar} (electron and hole) nature of transport 
in inhomogeneously doped semiconductors:
a spin-polarized {\it p-n} junction\cite{dassarma00,zutic01a,zutic01b} and 
a magnetic {\it p-n} junction.\cite{zutic02,zutic02b} A spin-polarized 
{\it p-n} junction is a {\it p-n} junction with a 
source spin injected externally into one or both regions ($p$ and $n$). 
The source spin can be injected either optically or electrically. 
We have demonstrated that 
nonequilibrium spin can be injected (transfered)
very effectively across the depletion layer (space-charge region), 
from both regions: by the majority carriers
into the respective minority region, and, {\it vice versa}, by the minority 
carriers into the respective majority region. Spin injection (throughout the
paper ``spin injection'' will mean spin injection through
the depletion layer, while externally injected spin will be referred to as
``source'' spin) by the minority carriers leads to spin accumulation in the majority region, 
with an effect of amplifying the spin 
and significantly extending the spin diffusion/drift length.\cite{zutic01a}
We have also shown that nonequilibrium spin can be stored 
and manipulated in a spin-polarized {\it p-n} junction by external bias--a 
spin capacitance effect.\cite{zutic01a}
Furthermore, a spin-polarized {\it p-n} junction can generate 
spin-polarized currents as a spin solar cell:\cite{zutic01b} 
when illuminated by circularly polarized light, a spin-polarized 
current flows in a {\it p-n} junction.

Magnetic {\it p-n} junctions\cite{zutic02} offer even more 
functionality by coupling equilibrium magnetism and nonequilibrium 
spin. A magnetic {\it p-n}
junction is formed by doping a {\it p-n} junction with magnetic
impurities, differently in the $p$ and $n$ regions. Magnetic impurities
induce large $g$ factors of the mobile carriers, thus the application of
a magnetic field results in a significant spin splitting of the carrier
bands.\cite{dietl94} If the doping is so large as to induce a ferromagnetic
order, the splitting appears also without magnetic field. 
The important question, of whether spin can be injected by 
the majority carriers from the magnetic majority region into the
nonmagnetic minority one, was answered negative. We have demonstrated 
that only if nonequilibrium spin is generated first in the
majority region, 
it can subsequently be injected through the depletion
layer. Spin can be also injected through the depletion layer
at large biases, since then, without any external spin source,
nonequilibrium spin is generated by the strong electric
field in the bulk regions. We have also shown that magnetoresistance
of a magnetic {\it p-n} junction increases exponentially 
with increasing magnetic field (that is, spin band splitting) at large
fields. Magnetic {\it p-n} junctions exhibit even giant 
magnetoresistance, when source spin is injected into the
majority region.  We have also predicted a spin-voltaic effect (the phenomenon related
to the Silsbee-Johnson spin-charge coupling\cite{silsbee80,johnson85}) where charge
current (or voltage in an open circuit) arises solely due to 
a nonequilibrium spin maintained in proximity to the magnetic region. 
Magnetic {\it p-n} junctions can also serve as spin valves,
since the direction of the zero-bias current can be reversed  
by reversing either the polarization of the source spin or the 
direction of the applied magnetic field.  

We have studied spin-polarized and magnetic {\it p-n} junctions
mainly numerically,\cite{zutic01a,zutic01b,zutic02} by solving a self-consistent set of 
recombination-relaxation and drift-diffusion equations, and Poisson's
equation. We have obtained solutions for the carrier and spin densities
and currents for small and large biases, and different values of magnetic
fields and the externally injected spin polarization. Numerical solution is indispensable
at large biases (large injection), where analytical methods are not 
available. Large bias solutions describe carrier and spin transport 
as {\it both} drift and diffusion, since drift currents due to 
electric fields are significant even outside of the depletion layer. 
The low injection regime is tractable analytically.
In Ref.~\onlinecite{zutic02} we have introduced a
heuristic analytical model which accounts well for the numerical 
findings, and explains all the important qualitative features of 
magnetic {\it p-n} junctions. In fact, our numerical
solutions show that the most interesting and potentially important
properties of magnetic junctions are at small biases; large biases
may still be useful for injecting spin across the depletion layer, 
or extracting spin from the bulk regions,\cite{zutic02} as described
in Sec.~\ref{sec:biasing}.) 

In this paper we formulate a general model of magnetic {\it 
p-n} junctions (the model includes spin-polarized {\it p-n} nonmagnetic
junctions as a particular case), following the classic formulation of 
Shockley of ordinary bipolar junctions.\cite{shockley50,tiwari92} 
The model describes magnetic {\it p-n} junctions at small biases (low injection), 
with arbitrary external (source) spin injection and band spin splitting
(magnetic field), within the limits of nondegenerate carrier
statistics. The paper has the dual role of describing the fundamental
properties of spin-polarized transport in inhomogeneous magnetic
semiconductors, while presenting a model simulation, based on the
recombination and relaxation limited bipolar drift and diffusion, 
of novel microelectonics spintronic devices. 
If semiconductor spintronics is to become a reality, then detailed transport
analyses of the type presented in this paper are essential.
The fully analytic nature of our theory makes our model simulation
particularly useful.

The paper is organized as follows. Section \ref{sec:model} introduces 
the model and  formulates its assumptions and approximations. 
Section \ref{sec:profile} describes the spatial profiles of the carrier 
and spin densities in the bulk regions, gives the boundary 
conditions for the densities, and discusses the I-V characteristics of magnetic 
{\it p-n} junctions. In Sec.~\ref{sec:discussion} we apply our theory
to several cases of interest: spin injection--through the depletion layer--by 
the majority carriers, spin pumping and spin amplification by the minority carriers, 
source spin injection by the biasing electrode, spin injection and
extraction at large biases, and magnetic drift effects in the carrier and
spin transport. 
Finally, we summarize our findings in 
Sec.~\ref{sec:summary}, where we also outline strategies for applying our theory to 
more realistic materials structures and more complex spintronic devices
based on magnetically inhomogeneous semiconductors.

\section{\label{sec:model}model}

\begin{figure}
\centerline{\psfig{file=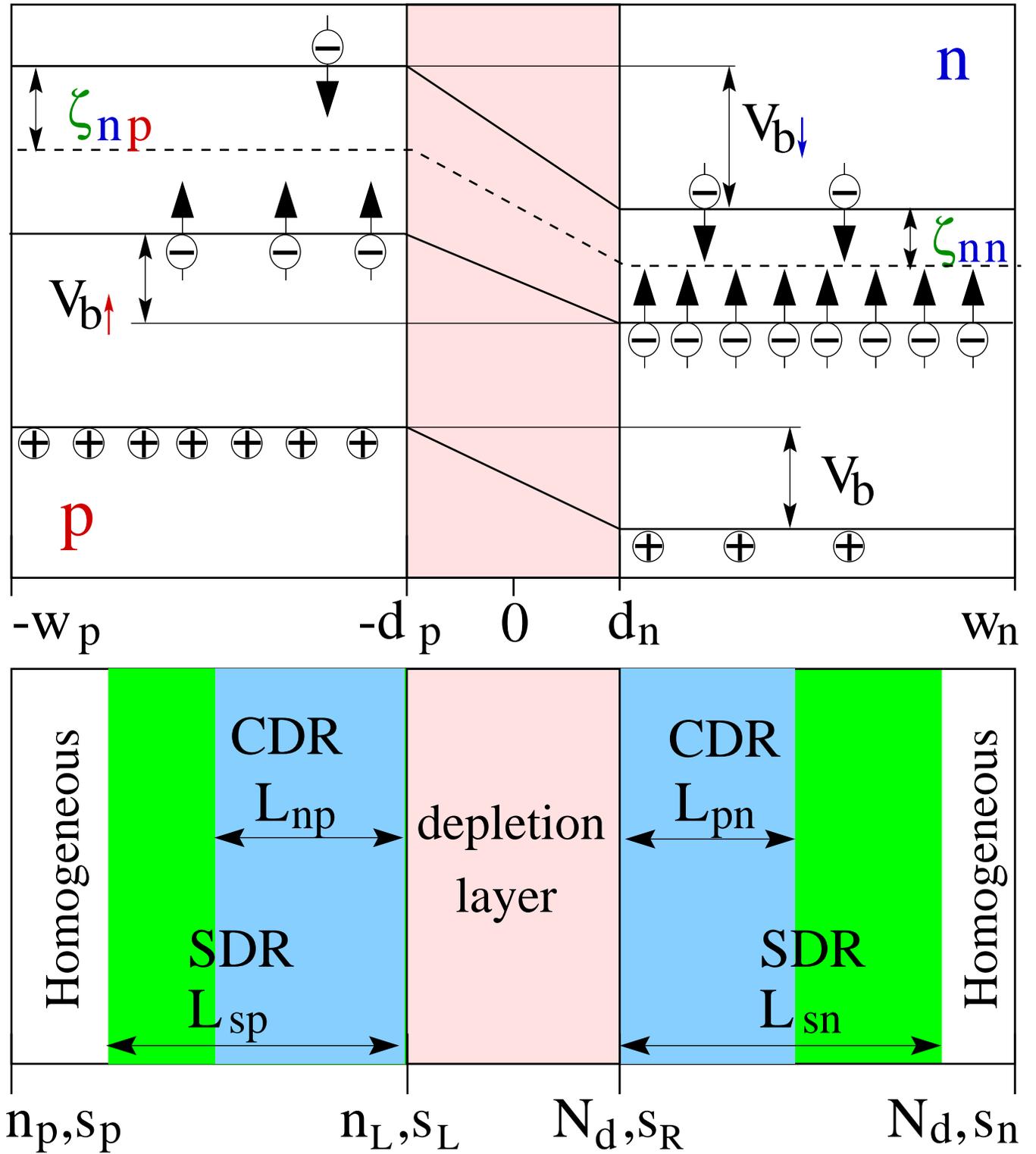,width=1\linewidth,angle=0}}
\caption{Schematics of a magnetic {\it p-n} junction. The
junction is $p$-doped from $-w_p$ to 0 and
$n$-doped from 0 to $w_n$. The depletion layer 
(space-charge region) 
forms at
$-d_p < x < d_n$. The upper figure depicts an inhomogeneously 
spin-split 
conduction band and a valence band without the spin splitting.
The conduction band spin splitting in the $n$ region is $2\zeta_{nn}$,
in the $p$ region it is $2\zeta_{np}$. The greater the $\zeta$ is, the more is the
lower band (here called the spin up sub-band)  populated. The
intrinsic built-in field across the depletion layer is $V_b$. 
For electrons the built-in field becomes explicitly spin dependent:
$V_{b\uparrow}=V_b+\zeta_{nn}-\zeta_{np}$ and  
$V_{b\downarrow}=V_b-\zeta_{nn}+\zeta_{np}$. 
The lower figure depicts 
regions with distinct transport characteristics: CDR are the 
(minority) carrier diffusion regions and SDR are the spin 
(here only electron) diffusion regions. The characteristic 
sizes of the regions are given by the corresponding diffusion
lengths, as indicated. The unshaded areas are
the homogeneous regions, where carrier and spin densities
assume their equilibrium values. The known (input) densities
of the model are $n_p$, $s_p$ at $-w_p$, and $N_d$, $s_n$ at $w_n$, 
while the densities at the depletion layer, $n_L$ and $s_L$ 
on the left side and $n_R=N_d$, $s_R$ on the right side, are calculated 
in the text. 
}
\label{fig:1}
\end{figure}

The basis for our model is a semiconductor {\it p-n} junction in which
carrier bands are inhomogeneously spin split: there is a finite
equilibrium spin polarization of the carriers, different
in the $p$ and $n$ regions.\cite{note} Large (comparable to the
thermal energy) spin splitting of carrier
bands can arise as a result of doping with magnetic impurities (which may,
but need not, contribute to the carrier densities). Magnetic impurities can
significantly increase the carrier $g$ factors (usually up to $g\approx 200$\cite{dietl94}), 
so that the application of a magnetic field $B$
induces large spin Zeeman splitting, $2\zeta=g\mu_B B$, of the bands ($\mu_B$ is
the Bohr magneton). Inhomogeneous
spin splitting can be realized either by inhomogeneous magnetic
doping in a homogeneous magnetic field, or by a homogeneous magnetic 
doping in an inhomogeneous magnetic field, or both.  
Our model applies equally well to ferromagnetic {\it p-n} junctions,
where bands are spin split even at zero magnetic field.
To keep the discussion transparent and to avoid complex notation, 
we consider only the conduction band to be spin split (that is, only 
electrons to be spin polarized), keeping holes unpolarized. 
This simplification does not affect our
conclusions, as electron and hole transports are fully
separated in our model. (Spin polarization of holes 
is treated in Appendices \ref{appendix:1} and \ref{appendix:2}.) 
The layout of a magnetic {\it p-n}
junction is shown in Fig.~\ref{fig:1}. The semiconductor is $p$-doped with $N_a$
acceptors (per unit volume) along the $x$ axis from $-w_p$ to 0, 
and $n$-doped with $N_d$ donors from 0 to $w_n$. The depletion
layer forms at $(-d_p,d_n)$. We are not concerned with the transition region
itself--we simply assume that it is steep enough (in fact, that it changes
over a region smaller than the Debye screening length) to support space 
charge, and that all the spin splitting 
changes occur only within the transition region, being constant 
in both $p$ and $n$ regions (the case of magnetic drift where the
splitting is inhomogeneous also in the bulk regions is treated in 
Sec.~\ref{sec:drift}).
 
We denote the electron density as $n=n(x)$ 
and the hole density as $p=p(x)$. The corresponding 
equilibrium values are $n_0$ and $p_0$,
and the deviations from the equilibrium values are $\delta n=n-n_0$ 
and $\delta p=p-p_0$. Electron spin density $s=s(x)$ 
(in equilibrium $s_0$ and deviation $\delta s=s-s_0$)
is a difference between the densities of spin up and spin
down electrons: $s=n_{\uparrow}-n_{\downarrow}$. As a measure
of spin polarization we use the spin polarization of the 
carrier density ({\it not} current): $\alpha=s/n$ 
(in equilibrium $\alpha_0$ and deviation $\delta \alpha=\alpha-\alpha_0$).
The equilibrium properties of magnetic {\it p-n} junctions
are discussed in Appendix \ref{appendix:1}, where
$n_0$, $p_0$, $s_0$, and the built-in potential  $V_b$
are calculated. 
The transport parameters of the carriers are diffusivities
$D_{nn}$ and $D_{np}$ of electrons in the $n$
and $p$ regions, electron lifetime $\tau_{np}$ in the 
$p$ region, and electron spin lifetime $T_{1p}$ and
$T_{1n}$ in the $p$ and $n$ regions. The unpolarized holes
are characterized by $D_{pn}$ and $\tau_{pn}$,
diffusivity and lifetime in the $n$ region. 
Throughout the paper, unless explicitly specified
otherwise, a single subscript denotes
the region or boundary ($p$, $n$, $L$, or $R$), while 
a double subscript denotes first the carrier type
or spin ($p$, $n$, or $s$) and then the 
region or the boundary (for example, $\tau_{pn}$ is the lifetime
of holes in the $n$ region). Terms
``majority'' (``minority'') will refer to electrons 
in the $n$ ($p$) region, and similarly for holes,
and {\it not} to the more (less) populated spin states, 
as is usual in the physics of magnetotransport. 
Similarly, the term bipolar bears no relation to 
spin, describing only the transport
carried by both electrons and holes. Finally,
terms ``bulk'' and, equivalently, ``neutral'' will  denote the 
regions outside the depletion layer, where, at 
low biases, charge neutrality is maintained. The notation is
summarized in Table \ref{tab:1}.

The junction is driven off equilibrium by applying bias and
injecting source spin. We place contact electrodes at $x=-w_p$
and $x=w_n$. We keep the left electrode general, capable of
injecting electrons, $\delta n_p\equiv \delta n(-w_p)\ne 0$, and spin,
$\delta s_p\equiv \delta s(-w_p)\ne 0$. This boundary condition 
covers magnetic diodes
(Ohmic contact, $\delta n_p=0$), and magnetic solar
cells and junction transistors ($\delta n_p\ne 0$). The right 
electrode is assumed to be Ohmic, $\delta p_n\equiv \delta p(w_n)=0$, but
able to inject spin, $\delta s_n\equiv \delta s(w_n)\ne 0$. The 
majority carriers in both regions are assumed constant:
$p=N_a$ in the $p$ side and $n=N_d$ in the $n$ side.
The  source spin injection, here considered to take place geometrically
at the contacts, can be realized   
either by the contact electrodes 
themselves (if the electrodes are magnetic), by optical orientation close
to the contact, 
or by electrical spin injection from a third electrode 
(say, transverse to the junction current). 
Different cases mean different boundary conditions for spin. 
For now we assume a third terminal injection
so that  $\delta s_p$ and $\delta s_n$ are free
parameters of the model; we will later, in Sec.~\ref{sec:contact} consider the
case of the contact 
(biasing) electrode source spin injection, where $\delta s_n$ will depend on 
the charge current in the junction.

To reduce the initial  drift and diffusion transport 
problem to a simple diffusion problem  in the neutral regions 
we need to know the boundary
conditions for the bulk regions at the depletion layer,
that is, the carrier and spin densities $n_L\equiv
n(-d_p)$, $s_L\equiv s(-d_p)$ at the left ($L$) and
$n_R\equiv n(d_n)$, $s_R\equiv s(d_n)$ at the right ($R$)
boundary of the depletion layer.
We will calculate these boundary densities in the subsequent sections.

We use several approximations to solve our model. First, we consider only low biases,
meaning that the applied forward voltage $V$
is smaller than the built-in field $V_b$, which is typically about 1 eV. 
At small biases the densities of the minority carriers are much smaller than  
the densities of the corresponding majority carriers (the small injection limit),
the electric field is 
confined to the depletion layer, and the bulk regions can be
considered neutral. We next assume that the temperature 
is large enough for the donors and 
acceptors to be fully ionized, so that $n=N_d$ 
and $p=N_a$ in the respective majority regions, and 
the carriers obey the nondegenerate 
Boltzmann statistics (limiting doping densities  to
about 10$^{18}$/cm$^3$ for typical semiconductors at room 
temperature).  
Finally, we consider only 
moderate spin splittings (much smaller than the built-in field), 
perhaps no greater than $5 k_B T$, since greater splittings 
can severely affect the band structure, and reduce the effective band gap.  

We have also made simplifying assumptions as to the 
band structure of the magnetic semiconductor. First, 
we neglect possible orbital degeneracy of the bands, and treat the
spin states as spin doublets. We also neglect the effects of
magnetic doping on the band structure (that is, changes in
$n_i$, additional band offsets, etc.) and that of the carrier
density on the band spin splitting.  The latter can be important 
in ferromagnetic semiconductors. However,
since it is the minority carriers which determine 
the transport across {\it p-n} junctions, it is unlikely
that a variation in the carrier density would 
appreciably affect our conclusions. We also 
assume that momentum and energy relaxation proceeds
much faster than carrier recombination and spin
relaxation, so that nonequilibrium, spin-dependent chemical potentials 
describe well the junction under an applied bias and with a source spin. 
Finally, we do not consider orbital effects due to the applied magnetic
field, although these can be included in our theory simply by allowing for a
magnetic dependence of diffusivities.

\begin{table}
\begin{tabular}{ll}
\hline
\hline
$N_d=$ & donor density \\
$N_a=$ & acceptor density \\
$n=$ & electron density (equilibrium $n_0$)\\
$p=$ & hole density ($p_0$)\\
$s=$ &  spin density ($s_0$)\\
$\alpha=$ & $s/n$,  spin polarization ($\alpha_0$)\\
$\delta \tilde{s}=$ & $s - \alpha_0 n $, the effective nonequilibrium spin density  \\
$J_n=$ &  electron particle current \\
$j_n=$ &  $-qJ_n$, electron charge current \\
$J_p=$ &  hole particle current \\
$j_p=$ &  $qJ_p$, hole charge current \\
$J_s=$ &  spin current \\
$n_p=$ &  electron density at $x=-w_p$ ($n_{0p}$)\\
$n_L=$ &  electron density at $x=-d_p$ ($n_{0L}=n_{0p}$)\\
$s_p=$ &  spin density at $x=-w_p$ ($s_{0p}$)\\
$s_n=$ &  spin density at $x=w_n$ ($s_{0n}$)\\
$s_L=$ &  spin density at $x=-d_p$ ($s_{0L}=s_{0p}$)\\
$s_R=$ &  spin density at $x=d_n$ ($s_{0R}=s_{0n}$) \\
$\tilde{w}_p=$ & $w_p-d_p$, effective width of the $p$ region \\
$\tilde{w}_n=$ & $w_n-d_n$, effective width of the $n$ region \\
$D_{nn}=$ & electron diffusivity in the $n$ region\\
$D_{np}=$ & electron diffusivity in the $p$ region\\
$\tau_{np}=$ & lifetime of electrons in the $p$ region\\
$\tau_{pn}=$ & lifetime of holes in the $n$ region\\
$L_{np}=$ & $\sqrt{D_{np}\tau_{np}}$, electron diffusion length in the $p$ region \\
$L_{pn}=$ & $\sqrt{D_{pn}\tau_{pn}}$, hole diffusion length in the $n$ region\\
$T_{1p}=$ & intrinsic spin lifetime in the $p$ region\\
$L_{1p}=$ & $\sqrt{D_{np}T_{1p}} $ intrinsic spin decay length in the $p$ region\\
$1/\tau_{sp}=$ & $1/\tau_{np}+1/T_{1p}$, spin decay rate in the $p$ region\\
$L_{sp}=$ & $\sqrt{D_{np}\tau_{sp}}$, spin diffusion length in the $p$ region\\
$T_{1n}=$ & spin lifetime in the $n$ region\\   
$L_{sn}=$ & $\sqrt{D_{nn}T_{1n}}$, spin diffusion length in the $n$ region\\
$V_b=$ & built-in potential \\
$V=$ & applied bias \\
\hline
\hline
\end{tabular}
\caption{Summary of the notation used in the text and in
Table \ref{tab:2}. All the spin parameters (spin density, spin lifetime, etc.)
relate to electrons. In the brackets are the equilibrium densities.  
}
\label{tab:1}
\end{table}

\section{\label{sec:profile}
carrier and spin transport in the neutral regions}

The transport of carriers and spin in magnetic {\it p-n}
junctions can be realistically described as drift and diffusion,
limited by carrier recombination and spin relaxation. The transport
equations were introduced in Ref.~\onlinecite{zutic02}, and have 
been solved numerically for a few important cases in 
Refs.~\onlinecite{zutic01a,zutic01b,zutic02}. 
Denoting the carrier (here electron) and spin currents as $J_n$ and $J_s$, 
the drift-diffusion equations are
\begin{eqnarray}\label{eq:Jn}
J_n&= &D_n(n \phi_t' + s\zeta' - n'), \\
\label{eq:Js}
J_s&=&D_n(s \phi_t'+ n\zeta' - s').
\end{eqnarray}
Here $\phi_t$ is the total local electrostatic potential, 
comprising both the built-in potential $\phi_b$  and applied bias
$V$ (the electric field is $E=-\phi_t'$), 
and
the magnetic drift is proportional to the spatial changes in the
band spin splitting, $\zeta'$ (see Fig.~\ref{fig:1}). Throughout this paper we express the
potentials and the energies in the units of $k_B T/q$ and $k_B T$,
respectively ($k_B$ is the Boltzmann constant, $T$ is the temperature, and 
$q$ is the proton  
charge). In a steady state  
carrier recombination and spin relaxation processes can be expressed  
through the  continuity equations for electrons and spin:
\begin{eqnarray} \label{eq:rec}
J_n'&=&-r(np-n_0p_0), \\ \label{eq:relax}
J_s'&=&-r(sp-s_0p_0)-\frac{\delta\tilde s}{T_{1}}, 
\end{eqnarray}
where $r$ is the electron-hole recombination rate and
$\tilde s\equiv s- \alpha_0 n$,   
expressing the fact that intrinsic
spin relaxation processes (spin-flip scattering, say,  
by phonons or impurities) conserve the local carrier density.\cite{optical84,zutic02}
Electron-hole
recombination also degrades spin, the fact reflected in the 
first term of Eq.~(\ref{eq:relax}). Equations (\ref{eq:Jn}),(\ref{eq:Js}),
(\ref{eq:rec}), and (\ref{eq:relax}), together with Poisson's equation
$\phi_t''=-\rho(q/\epsilon k_B T)$, where $\rho$ is the local charge density
and $\epsilon$ is the semiconductor's dielectric constant,
fully describe the steady-state carrier and spin transport  
in inhomogeneous magnetic semiconductors.\cite{zutic02} 
In the rest of the paper (except for Sec.~\ref{sec:drift}), 
magnetic drift force will play no explicit role, 
since we assume that the magnetic doping is uniform in the bulk regions. The 
inhomogeneity in the spin splitting, which is confined to the depletion region, will appear only 
through the boundary conditions.

At low biases, the case most important for device applications,
the problem of the carrier and spin transport in magnetic {\it p-n} junctions reduces
to the problem of carrier and spin diffusion in the neutral
regions.\cite{shockley50,tiwari92} This observation, to be useful, needs to be furnished 
with the boundary conditions for the carrier and spin densities at the depletion layer 
boundary ($n_L$, $s_L$, $n_R$, and $s_R$). Shockley's
model\cite{shockley50} evaluates the carrier densities in unpolarized {\it p-n}
junctions from the assumption that a quasiequilibrium is maintained
in the depletion layer even at applied (low) biases. This assumption alone
is insufficient to obtain both carrier and spin densities in a 
spin-polarized magnetic junction. We use, in addition, the continuity 
of the spin current in the depletion layer to calculate
the densities. A simple version of this model was introduced 
in Ref.~\onlinecite{zutic02}, where it was assumed that
(1) at a forward bias and with a source spin injected into the majority
region ($\delta s_n\ne 0$) the spin current at the depletion
layer, $J_{sR}$, vanishes, and (2) at a reverse bias, 
and with spin injected into the minority region ($\delta s_p\ne 0$),
all the spin entering the depletion region is 
swept by the large built-in field to the majority side.
Assumption (1) explains spin injection
of nonequilibrium spin through the depletion layer, while
(2) explains spin pumping by the minority carriers. Both 
assumptions will follow as special cases of the spin 
current continuity, in our model. 

In analogy with unpolarized {\it p-n} junctions,\cite{ashcroft76} there
are several regions with distinct transport characteristics
in spin-polarized magnetic {\it p-n} junctions, as illustrated in 
Fig.~\ref{fig:1}: (i) the depletion layer with space charge and large
carrier and spin drift and diffusion; (ii) the carrier
diffusion regions (CDR) which are neutral and where the
minority carriers' drift can be neglected. CDR are
characterized by carrier diffusion lengths
$L_{np}$ for electrons on the
$p$ side and $L_{pn}$ for holes
on the $n$ side; (iii) the spin diffusion regions (SDR),
which are neutral and where spin (both majority and
minority) drift can be neglected. SDR are characterized
by spin diffusion length
$L_{sp}$  on the $p$ side and $L_{sn}$ on the $n$ side; (iv) 
the homogeneous regions in the rest of the junction,
which are neutral, and where the carrier and spin densities
assume their equilibrium values. There is no diffusion,
only the majority carriers' drift.

This section presents a unified picture of carrier and
spin transport in magnetic {\it p-n} junctions. We first describe
the profiles of carrier and spin densities inside the bulk regions, 
as dependent on the densities at the depletion layer,  which
are calculated next by modifying Shockley's model to the spin
polarized case. The four (not independent) important assumptions 
used are (a) neutrality of the bulk regions, (b) small injection of the carriers
across the depletion layer and at the biasing contacts, 
(c) the existence of a thermal quasi-equilibrium
across the depletion layer even under applied bias and source spin,
and (d) continuity of spin current across the 
depletion layer. Our analytical results, summarized in Table \ref{tab:2},
show how the carrier density is influenced by both bias (as in the unpolarized
case) and nonequilibrium spin, and, {\it vice versa}, 
how nonequilibrium spin is influenced by both bias and nonequilibrium 
carrier density. This interplay is imprinted most significantly
in the dependence of the I-V characteristics of the magnetic
diodes on nonequilibrium spin. 

\subsection{Carrier and spin profiles}

\subsubsection{\label{sec:p} $p$ region}
In the $p$ region the hole density is uniform, $p=N_a$. 
Electrons are the minority carriers whose diffusion 
is governed by the equation
\begin{equation} \label{eq:np}
\delta n''=\frac{\delta n}{L_{np}^2},
\end{equation}
where the electron diffusion length is $L_{np}=\sqrt{D_{np}\tau_{np}}$.
We remind that if two subscripts are used in a label, the first denotes the 
carrier type ($p$ or $n$) or spin ($s$), and the second the region or
the boundary ($p$, $n$, $L$, or $R$); if only one subscript
is used, it denotes the region or the boundary. Equation (\ref{eq:np})
is obtained by combining Eqs.~(\ref{eq:Jn}) and
(\ref{eq:rec}), neglecting the electric drift force (magnetic drift vanishes
in the bulk regions),
and defining $1/\tau_{np}\equiv rN_a$. The boundary conditions for
the electron density are $\delta n_p$ at $x=-w_p$ and $\delta n_L$ 
(yet unknown) at $x=-d_p$. The boundary position of the depletion layer is not
fixed, but changes with the applied voltage and the equilibrium 
magnetization (through $V_b$, see Appendix \ref{appendix:1}) as
\begin{equation}
d_p=\sqrt{\frac{2\epsilon}{q}\frac{N_d}{N_a}\frac{V_b-V}{N_a+N_d}}.  
\end{equation}
It is useful to introduce $\tilde{w}_p=w_p-d_p$ to describe the effective width 
of the $p$ region. The solution of Eq.~(\ref{eq:np}) can then be written as
\begin{equation}
\delta n=\delta n_L \cosh(\eta_{np})+F_{np}\sinh(\eta_{np}),
\end{equation}
where $\eta_{np}\equiv (x+d_p)/L_{np}$ and 
\begin{equation}
F_{np}=\frac{\delta n_L\cosh(\tilde{w}_p/L_{np})-
\delta n_p}{\sinh(\tilde{w}_p/L_{np})}.
\end{equation}
``Flux'' parameters $F$ are central to our analysis, since they determine
the currents at the depletion layer. Effectively, $F$ measures 
the change in 
the nonequilibrium (here carrier $\delta n$) density over the length scales
of the (here carrier $L_{np}$) diffusion length:  
For a short $p$ region, $L_{np}\gg \tilde{w}_p$,
$F_{np}\approx (\delta n_L-\delta n_p)L_{np}/\tilde{w}_p$, while for
$L_{np}\ll \tilde{w}_p$, $F_{np}\approx \delta n_L$. 
The electron current profile, $J_n=-D_{np}\delta n'$, is
\begin{equation}
J_n=-\frac{D_{np}}{L_{np}}\left[\delta n_L \sinh(\eta_{np})+ 
F_{np}\cosh(\eta_{np})\right].
\end{equation}
At the depletion layer, $x=-d_p$, the current is 
\begin{equation} \label{eq:JnL}
J_{nL}=-\frac{D_{np}}{L_{np}}F_{np}.
\end{equation}

The spin density is also described by a diffusion equation. From 
Eqs.~(\ref{eq:Js}) and (\ref{eq:relax}), under the conditions of
charge 
neutrality and magnetic uniformity, we obtain
\begin{equation}\label{eq:sdifp}
\delta s''=\frac{\delta s}{L_{sp}^2}-\alpha_{0p}\frac{\delta n}{L_{1p}^2},
\end{equation}
where $L_{1p}=\sqrt{D_{np}T_{1p}}$ and the effective spin diffusion length in
the $p$ region is $L_{sp}=\sqrt{D_{np}\tau_{sp}}$, where
$1/\tau_{sp}=1/\tau_{np}+1/T_{1p}$ is the effective spin relaxation
rate, reflecting the fact that, in addition to intrinsic spin relaxation
processes, carrier recombination degrades spin. The second term in the RHS
of Eq.~(\ref{eq:sdifp}) acts as a local spin source, and appears because a change
in the electron density, $\delta n$, drives spin by intrinsic spin
relaxation processes to $\alpha_{0p}\delta n$ [see Eq.~(\ref{eq:relax})], 
thereby preserving the equilibrium spin polarization, but not the spin itself.   
The boundary conditions for the spin density are $\delta s_p=\delta s(-w_p)$ and, yet unknown, 
$\delta s_L=\delta s(-d_p)$ . The solution of Eq.~(\ref{eq:sdifp}) is
\begin{equation} \label{eq:sp}
\delta s= \delta \tilde{s}_L\cosh(\eta_{sp})+F_{sp}\sinh(\eta_{sp})+
\alpha_{0p}\delta n,
\end{equation}
where $\eta_{sp}\equiv (x+d_p)/L_{sp}$, $\delta \tilde{s}_L=\delta s_L-
\alpha_{0L}\delta n_L$ is the effective nonequilibrium spin at $L$,
and 
\begin{equation}
F_{sp}=\frac{\delta \tilde{s}_L\cosh(\tilde{w}_p/L_{sp})
-\delta \tilde{s}_p}{\sinh(\tilde{w}_p/L_{sp})}
\end{equation}
is a normalized spin flux with $\delta \tilde{s}_p= \delta s_p-\alpha_{0p}\delta n_{p}$.
For a large spin diffusion length, $L_{sp}\gg \tilde{w}_p$,
$F_{sp}\approx (\delta \tilde{s}_L-\delta \tilde{s}_p)L_{sp}/\tilde{w}_p$, while for
$  L_{sp}\ll \tilde{w}_p$, $F_{sp}\approx \delta \tilde{s}_L$.
The first two terms in the RHS of Eq.~(\ref{eq:sp}) describe the deviation
of the spin density from $\alpha_{0p}n$, while the last term represents the
deviation $\alpha_{0p}\delta n$ which is solely due to intrinsic
spin relaxation ($T_1$) processes.  
The spin current, $J_{s}=-D_{np}\delta s'$, has the profile
\begin{equation}
J_{s}=-\frac{D_{np}}{L_{sp}}\left[\delta \tilde{s}_L\sinh(\eta_{sp})+
F_{sp}\cosh(\eta_{sp})\right]+\alpha_{0p}J_n.
\end{equation}  
The first two contributions describe the spin flow due to spatial variations 
in $\delta\tilde{s}$, while the last term represents the spin flow associated
with the spin-polarized electron current. 
Finally, at the depletion layer, $x=-d_p$, the spin current is 
\begin{equation} \label{eq:JsL}
J_{sL}=-\frac{D_{np}}{L_{sp}}F_{sp}+\alpha_{0L}J_{nL}. 
\end{equation}
The first term can be neglected if the spin polarization is close to its 
equilibrium value (which is typically the case at small biases and 
no source spin). The second term is important for 
spin extraction at large biases (see Sec.~\ref{sec:biasing}).

\subsubsection{\label{sec:n}$n$ region}

In the $n$ region only spin diffusion needs to be examined, as to a very
good approximation $n=N_d$ (charge neutrality actually requires that 
$n=N_d+\delta p$, where $\delta p$ is the deviation of the hole density 
from equilibrium; this gives a small contribution to spin
density $\delta s_R$, as is discussed in Sec. \ref{sec:biasing} and Appendix \ref{appendix:3}).
Electron spin diffusion is described by the equation (obtained from 
Eqs.~(\ref{eq:Js}) and (\ref{eq:relax}) neglecting electric and magnetic 
drifts and recombination processes, as $p\ll N_d$)  
\begin{equation}\label{eq:sdifn}
\delta s''=\frac{\delta s}{L_{sn}},
\end{equation}
where $L_{sn}=\sqrt{D_{nn}T_{1n}}$. 
We introduce $\tilde{w}_n=w_n-d_n$ as the 
effective width of the neutral region, with bias and equilibrium
spin polarization dependent depletion layer boundary 
\begin{equation}
d_n=\sqrt{\frac{2\epsilon}{q}\frac{N_a}{N_d}\frac{V_b-V}{N_a+N_d}}.
\end{equation}
The boundary conditions for the spin density are $\delta s_R=\delta s(w_n)$
and $\delta s_n=\delta s(d_n)$. The solution of 
Eq.~(\ref{eq:sdifn}) is
\begin{equation} \label{eq:dsn}
\delta s = \delta s_R \cosh(\eta_{sn})+F_{sn}\sinh(\eta_{sn}),
\end{equation}
where $\eta_{sn}\equiv (x-d_n)/L_{sn}$ and
\begin{equation}
F_{sn}=\frac{\delta s_n-\delta s_R \cosh(\tilde{w}_n/L_{sn})}
{\sinh(\tilde{w}_n/L_{sn})}.
\end{equation}
The normalized flux is $F_{sn}\approx (\delta s_n-\delta s_R)L_{sn}/\tilde{w}_n$
for a short $n$ region, $L_{sn}\gg \tilde{w}_n$, while $F_{sn}\approx
-\delta s_R$ when $L_{sn}\ll \tilde{w}_n$. The spin current, $J_{s}=-D_{nn}\delta s'$, is 
\begin{equation}\label{eq:Jsn}
J_{s}=-\frac{D_{nn}}{L_{sn}}\left [ \delta s_R \sinh(\eta_{sn}) + 
F_{sn} \cosh(\eta_{sn}) \right ].  
\end{equation}
Finally, at the depletion layer, $x=d_n$, the spin current is
\begin{equation}\label{eq:JsR}
J_{sR}=-\frac{D_{nn}}{L_{sn}}F_{sn}.
\end{equation}
The spin current at the
depletion layer boundary is solely the diffusion current due to 
a spatially inhomogeneous nonequilibrium spin in the region. Electrons
with just the equilibrium spin polarization will not contribute to 
spin flow within the model approximations (see Sec.~\ref{sec:biasing} 
for a discussion of how the neglected terms affect the carrier
and spin transport).

\subsection{Carrier and spin densities at the depletion layer}

Let $\phi(x)$ be the electrostatic potential resulting from the
application of applied bias $V$ (that is, not including the
equilibrium built-in potential $V_b$). We assume that all the
applied bias drops within the high resistance, carriers devoid, depletion layer:
\begin{equation}
\phi(d_n) - \phi(-d_p)=V,
\end{equation}
so that $\phi$ is constant in the bulk regions. 
Further, let $\mu$ be the deviation of the nonequilibrium chemical
potential from its equilibrium value; $ \mu$ is generally 
spin dependent: we will denote it as $ \mu_\uparrow$ for spin up and $ \mu_{\downarrow}$
for spin down electrons. That $ \mu$ is a
good description of the carrier and spin off-equilibrium energy distribution 
follows form the well established fact that energy and momentum
relaxation proceeds much faster than carrier recombination and spin relaxation. 
For a nondegenerate statistics, spin up and spin down electron densities
can be written as
\begin{eqnarray}
n_{\uparrow}(x)&=&n_{\uparrow 0}(x)\exp[\phi(x)+ \mu_{\uparrow}(x)], \\
n_{\downarrow}(x)&=&n_{\downarrow 0}(x)\exp[\phi(x)+ \mu_{\downarrow}(x)], 
\end{eqnarray} 
where $n_{\uparrow 0}$ and $n_{\downarrow 0}$ are the equilibrium
values; we have made explicit the fact that all the quantities describing
the densities vary in space. The electron, $n=n_\uparrow+n_\downarrow$,
and spin, $s=n_\uparrow-n_\downarrow$,  densities are
\begin{eqnarray}\label{eq:bcn}
n& =& \exp(\phi+ \mu_+)\left [n_0 \cosh( \mu_-)+
s_0\sinh( \mu_-)\right ], \\ \label{eq:bcs}
s& =& \exp(\phi+ \mu_+)\left[n_0 \sinh( \mu_-)+
s_0\cosh( \mu_-)\right ],
\end{eqnarray}   
where $ \mu_{\pm}\equiv 
( \mu_{\uparrow}\pm \mu_{\downarrow})/2$. Finally, the spin
polarization
\begin{eqnarray}
\alpha=\frac{\tanh( \mu_{-})+\alpha_0}{1+\alpha_0\tanh( \mu_{-})}
\end{eqnarray} 
depends on $ \mu_{-}$ only (while $n$ and $s$ depend on both
$ \mu_{+}$ and $ \mu_{-}$).

Substituting Eqs.~(\ref{eq:bcn}) and (\ref{eq:bcs}) into the equations
(\ref{eq:Jn}) and (\ref{eq:Js}) for the electron carrier and spin currents,
we obtain,
\begin{eqnarray} \label{eq:Jnmu}
J_n&=&-D_n\left (n  \mu_{+}' + s   \mu_{-}'\right ), \\
J_s&=&-D_n\left (n  \mu_{-}' + s  \mu_{+}'\right ).
\end{eqnarray}
It may be tempting to associate $\mu_{+}$ with only charge, and
$\mu_{-}$ with only spin (as done, for example, 
in Ref.~\onlinecite{dyakonov71}). It would then 
follow from Eq.~(\ref{eq:Jnmu}) that in a semiconductor with a uniform
carrier density a charge current would flow (or a spin emf would
appear) if a nonequilibrium spin gradient (or, equivalently here, spin 
polarization gradient) would be maintained.\cite{dyakonov71} This is wrong,
as can be seen directly from Eq.~(\ref{eq:Jn}) which shows that 
spin can contribute to charge current only through magnetic drift (see
Sec.~\ref{sec:drift}),
$\zeta'$. Although $ \mu_{-}$ indeed suffices to determine $\alpha$,
it also influences $n$. If $n$ is to be uniform and $ \mu_{-}$
has a finite gradient, then $ \mu_{+}$ must change to 
ensure that $n$ is unchanged, as follows from Eq.~(\ref{eq:bcn}).
However, a spin emf due to spin polarization gradient would appear in degenerate
semiconductors or metals,\cite{dyakonov71} as mobilities and diffusivities
for spin up and down species would generally be different in this 
case, and spin diffusion directly affects charge current.\cite{zutic02}

\subsubsection{\label{sec:shockley} Shockley's condition of constant chemical potentials}

We now apply the condition of constant chemical
potentials in the depletion layer to connect the charge and
spin densities at the left ($L$) and right ($R$) depletion layer edges. 
First notice that
\begin{equation}\label{eq:const}
\tanh( \mu_-)=\frac{\alpha(x)-\alpha_0(x)}{1-\alpha(x)\alpha_0(x)}
\equiv \rm const, 
\end{equation}
from which follows that the spin polarizations at $L$ and $R$ are 
connected without an explicit dependence on bias.
We will now express $\alpha_L$, $n_L$, and $s_L$ 
in terms of the nonequilibrium spin polarization in the $n$ region, 
$\delta \alpha_R=\delta s_R/N_d$; we will evaluate $\delta \alpha_R$ 
explicitly from the input parameters in the next section.  

It follows from Eq.~(\ref{eq:const}) that 
\begin{equation}\label{eq:aL}
\alpha_L=\frac{\alpha_{0L}(1-\alpha_{0R}^2)+\delta 
\alpha_R(1-\alpha_{0L}\alpha_{0R})}
{1-\alpha_{0R}^2+\delta \alpha_R(\alpha_{0L}-\alpha_{0R})}.
\end{equation}
If $\delta \alpha_R=0$, then $\alpha_L=\alpha_{0L}$. In 
other words, only {\it nonequilibrium} spin can be injected from
the majority region through the depletion layer. In the
case of a homogeneous spin splitting ($\alpha_{0L}=\alpha_{0R}$),
$\delta \alpha_L=\delta\alpha_{R}$, that is, the nonequilibrium 
spin polarization is constant across the depletion layer.
Also note that 
$\alpha_L$ depends on the applied bias only 
implicitly, through the possible bias dependence of $\delta\alpha_R$.

The carrier and spin densities at $L$ are determined by 
both $\delta \alpha_R$ and $V$. Equations (\ref{eq:bcn}) and 
(\ref{eq:bcs}) yield
\begin{eqnarray} \label{eq:nL}
n_L&=&n_{0L}e^V\left 
( 1+\delta \alpha_R \frac{\alpha_{0L}-\alpha_{0R}}{1-\alpha_{0R}^2}\right ), \\
\label{eq:sL}
s_L&=&s_{0L}e^V\left (1+\frac{\delta \alpha_R}{\alpha_{0L}}\frac{1-\alpha_{0L}\alpha_{0R}}
{1-\alpha_{0R}^2} \right ).
\end{eqnarray}
In the absence of nonequilibrium spin ($\delta \alpha_R=0$), the
above formulas reduce to the well known Shockley relation 
for the minority carrier density at the depletion layer,\cite{shockley50}
$n_L=n_{0L}\exp(V)$, and the analogous formula for spin,
$s_L=s_{0L}\exp(V)$, so that the equilibrium spin polarization
$\alpha_L=\alpha_{0L}$ is preserved. Equations (\ref{eq:nL})
and (\ref{eq:sL}) demonstrate the interplay between charge
and spin in magnetic {\it p-n} junctions: nonequilibrium 
spin $\delta \alpha_R$ can significantly affect the minority 
carrier density (thus the junction I-V characteristics, as
will be shown in Sec.~\ref{sec:IV}) and spin, while bias affects 
both the carrier and spin densities.
If the band spin splitting is homogeneous ($\alpha_{0L}=\alpha_{0R}$), 
nonequilibrium spin does not influence the minority
carrier density [and affects the spin density in a trivial
way: $s_L=n_{0L}\alpha_L\exp(V)$]. Equation (\ref{eq:nL})
suggests that the charge response, $\delta n_L$, to nonequilibrium
spin can be maximized by maximizing the difference in the
equilibrium spin polarizations, $| \alpha_{0L}-
\alpha_{0R}|$, and having $\alpha_{0R}$ as close
to $\pm 1$ as possible (the case of $\alpha_{0R}=\pm 1$
is pathological, and is excluded from our theory by the 
assumption of small injection, whereby $n_L \ll N_a, N_d$).

\subsubsection{\label{sec:continuity}Continuity of spin current in the depletion layer}

In the previous sections $\delta\alpha_R$ was treated as an
unknown input parameter to obtain the carrier and spin profiles,
and specifically the carrier and spin densities at $x=-d_p$.
Calculation of $\delta \alpha_R$ is performed in this section. The 
knowledge of $\delta \alpha_R$ will complete the formalism necessary to calculate 
any quantity of the magnetic {\it p-n} junction under
general conditions of applied bias and source spin, with the
stated constraints of the model. In the spin-equilibrium case 
($\delta \alpha=0$)
the calculation made in the
preceding section suffices to get all the necessary boundary conditions.
The reason is that the carrier density in the majority side is uniform,
$n=N_d$. Spin, however, does not behave similarly to the majority carriers
even in the majority region. Spin can be injected into the majority 
region, and diffuses, rather than drifts, there. This is why the unknown 
$\delta \alpha_R$ needs to be specified 
by another condition. Here we apply the condition of the continuity of spin 
current in the depletion layer. Physical justification for this
condition is the fact that in the depletion layer, devoid of carriers and spin, 
spin relaxation, proportional to the spin density, is inhibited. One can
write from Eq.~(\ref{eq:relax})
\begin{equation}\label{eq:bcsc}
J_{sR}=J_{sL}-J_{s,\rm relax},
\end{equation}
where $J_{s, \rm relax}$ is the spin relaxation
current (similar to the carrier recombination current used in treating
unpolarized junctions\cite{tiwari92}),
\begin{equation}\label{eq:bcsc1}
J_{s,\rm relax}=\int_{-d_p}^{d_n}dx\left [r(sp-s_0p_0)+\frac{s-\tilde{s}}{T_1}  
\right ].
\end{equation}
We neglect $J_{s,\rm relax}$ in the following treatment.\cite{neglect}

Equations (\ref{eq:JsL}), (\ref{eq:JsR}), (\ref{eq:nL}), and (\ref{eq:sL}), together
with Eq.~(\ref{eq:bcsc}), form a full, self-consistent set of equations needed to extract
$\delta s_R$ (or, equivalently, $\delta \alpha_R$), and thus complete the structure of 
the model. In the process of extracting $\delta s_R$, we apply the condition
of low injection, and neglect the terms of the order of $n_{0L}\exp(V)$ when compared
to $N_d$.  The result is
\begin{equation} \label{eq:dsR}
\delta s_R=\gamma_0\delta s_n+\gamma_1 \delta \tilde{s}_p
+\gamma_2\alpha_{0L}\delta n_p-\gamma_3s_{0L}\left (e^V-1\right ), 
\end{equation}
where the geometric/transport factors are
\begin{eqnarray}
\gamma_0&=&1/\cosh(\tilde{w}_n/L_{sn}), \\
\gamma_1&=&\left (\frac{D_{np}}{D_{nn}} \right )\left 
(\frac{L_{sn}}{L_{sp}} \right ) 
\frac{\tanh(\tilde{w}_n/L_{sn})}{\sinh(\tilde{w}_p/L_{sp})},\\ 
\gamma_2&=&\left (\frac{D_{np}}{D_{nn}} \right )\left 
(\frac{L_{sn}}{L_{np}} \right) 
\frac{\tanh(\tilde{w}_n/L_{sn})}{\sinh(\tilde{w}_p/L_{np})},\\ 
\gamma_3&=&\gamma_2\cosh(\tilde{w}_p/L_{np}).
\end{eqnarray}
Equation (\ref{eq:dsR}) expresses $\delta s_R$ in terms of the known input 
parameters, and can be used as an input for determining the carrier and spin densities
at the  depletion layer, as well as the carrier and spin profiles in the bulk regions.
The first contribution to $\delta s_R$ comes from the source spin at the
right contact, $\delta s_n$. The second and the third terms in the RHS of
Eq.~(\ref{eq:dsR}) come from the source spin and the carrier densities at the
left contact, and a  result of spin injection by the minority electrons
through the depletion layer. Finally, the last term, which usually 
is negligible, results from the spin flow of the minority 
electrons having the equilibrium spin polarization (that is, as if no spin
or minority electron source were present). This term, for large forward
biases, leads to spin extraction (see Sec.~\ref{sec:biasing}). In most 
practical cases the source spin is injected either in the majority 
or in the minority regions, not both. Then the contributions to $\delta s_R$ can 
be considered separately, with either the first, or the second and the
third terms in the RHS of Eq.~\ref{eq:dsR} contributing. The last term 
(that with $\gamma_3$) can
be usually neglected in the low injection limit.  
Implications of Eq.~(\ref{eq:dsR}) for spin-polarized transport in magnetic 
{\it p-n} junctions will be explored in Sec.~\ref{sec:discussion}.

The content of this and the previous sections is summarized in Table \ref{tab:2}.

\begin{table}
\begin{tabular}{|l|c|}
\hline
\multicolumn{2}{|c|}{$p$ region carrier density and current} \\
\hline\hline
\multicolumn{2}{|c|}{$\delta n''=\delta n/L_{np}^2$} \\
\multicolumn{2}{|c|}{$\delta n=\delta n_L \cosh(\eta_{np})+F_{np}\sinh(\eta_{np})$} \\
\hline
\multicolumn{2}{|c|}{$J_n=-D_{np}\delta n'$} \\
\multicolumn{2}{|c|}{$J_n=-(D_{np}/L_{np})[\delta n_L\sinh(\eta_{np})+F_{np}\cosh(\eta_{np})]$} \\
\hline 
$\eta_{np}$ & $(x+d_p)/L_{np}$ \\
$F_{np}$ & $ [\delta n_L\cosh(\tilde{w}_p/L_{np})-\delta n_p]/\sinh(\tilde{w}_p/L_{np})$ \\ 
$\delta n_L$ & 
$n_{0L}e^V\left [1+\delta \alpha_R (\alpha_{0L}-\alpha_{0R})/(1-\alpha_{0R}^2)\right]-n_{0L}$\\
$J_{nL}$ & $-(D_{np}/L_{np}) F_{np}$\\
\hline\hline
\multicolumn{2}{|c|}{$p$ region spin density and current} \\
\hline\hline
\multicolumn{2}{|c|}{$\delta s''=\delta s/L_{np}^2+\delta \tilde{s}/L_{1p}^2$} \\
\multicolumn{2}{|c|}{$\delta s=\delta \tilde{s}_L\cosh(\eta_{sp})+
F_{sp}\sinh(\eta_{sp})+\alpha_{0p}\delta n$} \\
\hline 
\multicolumn{2}{|c|}{$J_s=-D_{np}\delta s'$} \\
\multicolumn{2}{|c|}{$J_{s}=-(D_{np}/L_{sp})
\left [\delta \tilde{s}_L\sinh(\eta_{sp})+F_{sp}\cosh(\eta_{sp})\right ]+\alpha_{0p}J_n$} \\ 
\hline
 $\eta_{sp}$ & $(x+d_p)/L_{sp}$ \\
 $\delta \tilde{s}_L$ & $\delta s_L -\alpha_{0L}\delta n_L$ \\  
 $\delta \tilde{s}_p$ & $\delta s_p -\alpha_{0p}\delta n_p$ \\
 $F_{sp}$ & $ [\delta \tilde{s}_L\cosh(\tilde{w}_p/L_{sp})-
\delta \tilde{s}_p]/\sinh(\tilde{w}_p/L_{sp})$ \\
$\delta s_L$ & $s_{0L}
e^V\left [1+(\delta \alpha_R/\alpha_{0L})
(1-\alpha_{0L}\alpha_{0R})/(1-\alpha_{0R}^2)\right] -s_{0L}$\\
$J_{sL}$ & $ -(D_{np}/L_{sp})F_{sp}+\alpha_{0L}J_{nL}$ \\ 
\hline\hline
\multicolumn{2}{|c|}{$n$ region spin density and current} \\
\hline\hline
\multicolumn{2}{|c|}{$\delta s''=\delta s/L_{sn}^2$} \\
\multicolumn{2}{|c|}{$\delta s=\delta s_R\cosh(\eta_{sn})+
F_{sn}\sinh(\eta_{sn})$} \\
\hline
\multicolumn{2}{|c|}{$J_s=-D_{nn}\delta s'$} \\
\multicolumn{2}{|c|}{$J_s=-(D_{nn}/L_{sn})
\left [\delta s_R \sinh(\eta_{sn})+ F_{sn}\cosh(\eta_{sn})\right ]$} \\
\hline
$\eta_{sn}$ & $(x-d_n)/L_{sn}$ \\
$F_{sn}$ & $\left [\delta s_n -\delta s_R \cosh(\tilde{w}_n/L_{sn})
\right ]/ \sinh(\tilde{w}_n/L_{sn})$ \\
$\delta s_R$ & $\gamma_0\delta s_n+
\gamma_1 \delta \tilde {s}_p + \gamma_2 \alpha_{0L}\delta n_p  
-\gamma_3 s_{0L}\left (e^V-1\right )$ \\ 
$\gamma_0$& $1/\cosh(\tilde{w}_n/L_{sn})$ \\
$\gamma_1$ & $(D_{np}/D_{nn})(L_{sn}/L_{sp})
[\tanh(\tilde {w}_n/L_{sn})/\sinh(\tilde {w}_p/L_{sp})]$ \\
$\gamma_2$ & $(D_{np}/D_{nn})(L_{sn}/L_{np})
[\tanh(\tilde {w}_n/L_{sn})/\sinh(\tilde {w}_p/L_{np})]$ \\
$\gamma_3$& $\gamma_2\cosh(\tilde{w}_p/L_{np}) $ \\ 
$J_{sR}$ & $ -(D_{nn}/L_{sn})F_{sn} $ \\
\hline
\end{tabular}
\caption{The carrier and spin densities and currents in the bulk 
regions of a magnetic {\it p-n} junction. Only electrons are spin 
polarized (spin polarization of holes is treated in Appendices
\ref{appendix:1} and \ref{appendix:2}). 
For both the $p$ and $n$ regions, the diffusion 
equations and the equations for currents, as well as
the explicit formulas describing the spatial profiles of the densities
and currents in the bulk regions are given. The notation is 
summarized in Table \ref{tab:1}.
}
\label{tab:2}
\end{table}

\subsection{\label{sec:IV}I-V characteristics}

Charge current in a magnetic {\it p-n} junction is driven
by both external bias and source spin. Neglecting
carrier recombination in the depletion layer, the charge
electron current is the current that appears at the depletion
layer in the minority side, $x=-d_p$: $j_n=-qJ_{nL}$. 
Equation
(\ref{eq:JnL})  gives
\begin{equation}\label{eq:jn}
j_n=j_{0n}+j_{1n}+j_{2n},
\end{equation}
where 
\begin{eqnarray} \label{eq:j0n}
j_{0n}&=&j_{gn}\left (e^V-1\right ), \\ \label{eq:j1n}
j_{1n}&=&j_{gn}e^V \delta \alpha_R\frac{\alpha_{0L}-\alpha_{0R}}{1-\alpha_{0R}^2}, \\
\label{eq:j2n}
j_{2n}&=&j_{gn} \frac{1}{\cosh(\tilde{w}_p/L_{np})}\frac{\delta n_p}{n_{0L}}. 
\end{eqnarray}
By $j_{gn}$ we denote the electron generation current (current of thermally
excited electrons in the $p$ region close to the depletion layer\cite{tiwari92}):
\begin{equation}
j_{gn}=\frac{qD_{np}}{L_{np}}n_{0L}\coth\left ( \frac{\tilde{w}_p}{L_{np}}\right ).
\end{equation}
The generation current depends on the equilibrium magnetization
through $n_{0L}$ (see Appendix \ref{appendix:1}).
A magnetic {\it p-n} junction works as a diode when both electrodes
are Ohmic ($\delta n_p=0$), in which case $j_n=j_{0n}+j_{1n}$. This
current can be also written as  
\begin{equation} \label{eq:current}
j_{0n}+j_{1n}=j_{gn}\frac{\delta n_L}{n_{0L}},
\end{equation}
a notation which emphasizes the crucial role of the minority
carrier density at the depletion layer for charge transport.
Equation (\ref{eq:j0n}) describes the usual rectification current, which (for
an Ohmic contact) is the only 
carrier current in magnetically homogeneous junctions ($\alpha_{0L}=\alpha_{0R}$),
or in junctions lacking nonequilibrium spin ($\delta \alpha_R=0$). Once a nonequilibrium
spin is present, and the carrier bands are inhomogeneously spin split,
the current is modified by $j_{1n}$, the {\it spin-voltaic} current,  
the charge current caused by nonequilibrium spin. The spin-voltaic
current does not vanish at zero bias, giving rise to the spin-voltaic
and spin-valve effects\cite{zutic02} discussed in Sec.~\ref{sec:spinvoltaic}. 
Including the hole current (see Appendix \ref{appendix:2}),
the total charge current reads
\begin{equation} \label{eq:tc}
j=j_n+j_p.   
\end{equation}
Here we consider holes to be unpolarized, so that
\begin{equation}
j_p=j_{gp}(e^V-1),
\end{equation}
with 
\begin{equation}
j_{gp}=\frac{qD_{pn}}{L_{pn}}p_{0R}\coth(\tilde{w}_n/L_{pn})
\end{equation}
being the hole generation current. The hole current 
is affected by magnetic field only through  $p_{0R}$ (see Appendix
\ref{appendix:1}). If also holes would be spin polarized, the hole
current would depend on the nonequilibrium hole spin polarization, and
would exhibit all the spin phenomena we discuss for electrons. The 
corresponding formulas are presented in Appendix \ref{appendix:2}.

For spin injection problems it is often useful to consider the spin polarization
of the charge current, not only the density spin polarization $\alpha$. The
current spin polarization is defined as $\alpha_J=j_s/j$, where $j_s$ is the
spin current associated with charge flow. In our case of only electrons
being spin polarized, $j_s=-qJ_s$. Since $j$ is a conserved quantity,
the spin polarization profile is the same as the profile of the spin
current, already given in the previous sections. As will also be demonstrated
in the discussion of particular cases of interest, $\alpha_J$  can
differ significantly from $\alpha$. Unlike for $\alpha$, for example, the 
magnitude of $\alpha_J$ can be greater (even much greater) than unity 
(if spin up and down electrons flow in opposite directions). The knowledge
of the current spin polarization is essential particularly in studies of
spin injection, where typically one assumes that $\alpha_J$ is conserved
across the injection interface (see Sec.~\ref{sec:contact}), as a result
of the continuity of the spin current.

We close this section by explaining qualitatively the physics behind
the spin-voltaic current $j_{1n}$. Equation (\ref{eq:jn}) 
can be understood rather simply by considering
the balance between the recombination and generation currents in the
depletion layer.~\cite{ashcroft76} In the following we put $\delta n_p=0$, to simplify the discussion.
Let $\zeta_{nn}$ and $\zeta_{np}$ denote the conduction band splitting
in the $n$ and $p$ regions, respectively, as illustrated in Fig.~\ref{fig:1}. 
The recombination electron current is the current of the majority electrons 
flowing from $n$ to $p$. It is essentially the current of electrons with enough energy to cross
the potential barrier in the depletion layer. This barrier is different for
spin up ($V_{b\uparrow}=V_b+\zeta_{nn}-\zeta_{np}$) and spin down ($V_{b\downarrow}=V_b-\zeta_{nn}+
\zeta_{np}$) electrons. Within the Boltzmann
statistics the recombination current of spin up and down electrons, under 
applied bias $V$, is 
\begin{eqnarray}
j_{r \uparrow}=K n_{R\uparrow}e^{-V_b-\zeta_{nn}+\zeta_{np}+V}, \\
j_{r \downarrow}=K n_{R\downarrow}e^{-V_b+\zeta_{nn}-\zeta_{np}+V}.
\end{eqnarray}
where $K$ is a spin-independent constant. The recombination current is proportional
to the number of electrons $n_R$ available for thermal activation over the barrier, 
and the thermal activation Boltzmann factor $\exp(-V_b+V)$. 

The generation currents are the electron currents (flowing from $p$ to $n$) 
due to the minority electrons thermally generated in the diffusion region on
the $p$ side (Fig.~\ref{fig:1}), and swept by the large built-in field to 
the $n$ side. The generation currents are bias independent, and must equal the 
corresponding recombination currents if $V=0$, so that no net current flows 
in equilibrium. Thus
\begin{eqnarray}
j_{g \uparrow}=K n_{0R\uparrow}e^{-V_b-\zeta_n+\zeta_p}, \\
j_{g \downarrow}=K n_{0R\downarrow}e^{-V_b+\zeta_n-\zeta_p}.
\end{eqnarray}
The total electron charge current, $j_n=j_{r \uparrow}-j_{g \uparrow}+
j_{r \downarrow}-j_{g \uparrow}$, can be expressed
through the equilibrium and nonequilibrium electron spin polarizations, using
formulas from Appendix \ref{appendix:1}. The result is
\begin{equation}
j_n=Kn_{0L}\left [e^V\left (1+\frac{\alpha_{0L}-\alpha_{0R}}{1-
\alpha_{0R}^2}\right )-1  \right],
\end{equation}
which is, up to a constant, Eq.~(\ref{eq:jn}) (the
constant $K$, which is proportional to the generation current, can be obtained 
rigorously only by solving the corresponding diffusion equations).
The above reasoning explains the spin-voltaic effect in magnetic {\it p-n} junctions 
as resulting from the disturbances of the balance between the generation and recombination currents.
The nonequilibrium spin itself, $\delta \alpha_R$, which is an input for Eq.~(\ref{eq:j1n}),
must be obtained by considering the full set of assumptions leading to Eq.~(\ref{eq:dsR}).

\section{\label{sec:discussion}discussion}

As an application of our theory we discuss several important manifestations
of spin-polarized bipolar transport in magnetic {\it p-n} junctions, 
and illustrate the examples numerically with GaAs materials parameters.
The specific cases we consider are spin injection (through the
depletion layer) by the
majority carriers, spin pumping by the minority carriers,
the spin-voltaic effect, external (source) spin injection by the biasing
electrode, spin injection and extraction at large biases, and magnetic
drift in the neutral regions.

The reason for choosing GaAs for numerical examples is that GaAs
is the best studied semiconductor for spin properties.\cite{optical84}
Spin can be injected into GaAs both optically and electrically,
and high quality magnetic hybrid semiconductor structures
based on GaAs can be potentially fabricated, as underlined by the
discovery of ferromagnetic (Ga,Mn)As.~\cite{ohno98,esch97} For integration with semiconductor
technology, however, it would be much more desirable to have Si-based
spintronic devices. Although optical orientation~\cite{optical84} of electron spins
in Si is not effective because of the band structure (unlike GaAs, Si is not 
a direct band-gap semiconductor), there seems to be no fundamental 
reason why spin could not be injected into Si electrically;
thus far, however,  electrical spin injection into Si has proved 
elusive.~\cite{jia96}
In addition to the economic reasons of easy technological integration, Si could
offer other advantages over GaAs, such as (expected) longer spin
relaxation times (due to the weak spin-orbit coupling and the 
absence of the D'yakonov-Perel' mechanism~\cite{optical84,fabian99} 
of spin relaxation in centrally symmetric Si), and much larger 
intrinsic carrier density $n_i$ (important for bipolar conduction).

The numerical examples in the following sections are all based on a 
symmetric GaAs magnetic diode with the fixed parameters $N_a=N_d=10^{16}$ cm$^{-3}$,
$n_i=1.8\times 10^{6}$ cm$^{-3}$, $D_{n}=100$ cm$^{2}$/s, and $\tau_n=T_1=
1$ ns (equal in both regions), $w_p=w_n=3$ $\mu$m. The derived parameters
are $L_{np}\approx 3.2$ $\mu$m, $L_{sp}\approx 2.2$ $\mu$m, and $L_{sn}\approx 3.2$ $\mu$m.  
Other parameters (bias, equilibrium and
nonequilibrium spin) will be specified according to the physical
situation. The materials parameters are for  room temperature, so
the chemical potentials will be given in the units of $k_B T\approx 25$ meV.

\subsection{Spin injection by the majority carriers}

\begin{figure}
\centerline{\psfig{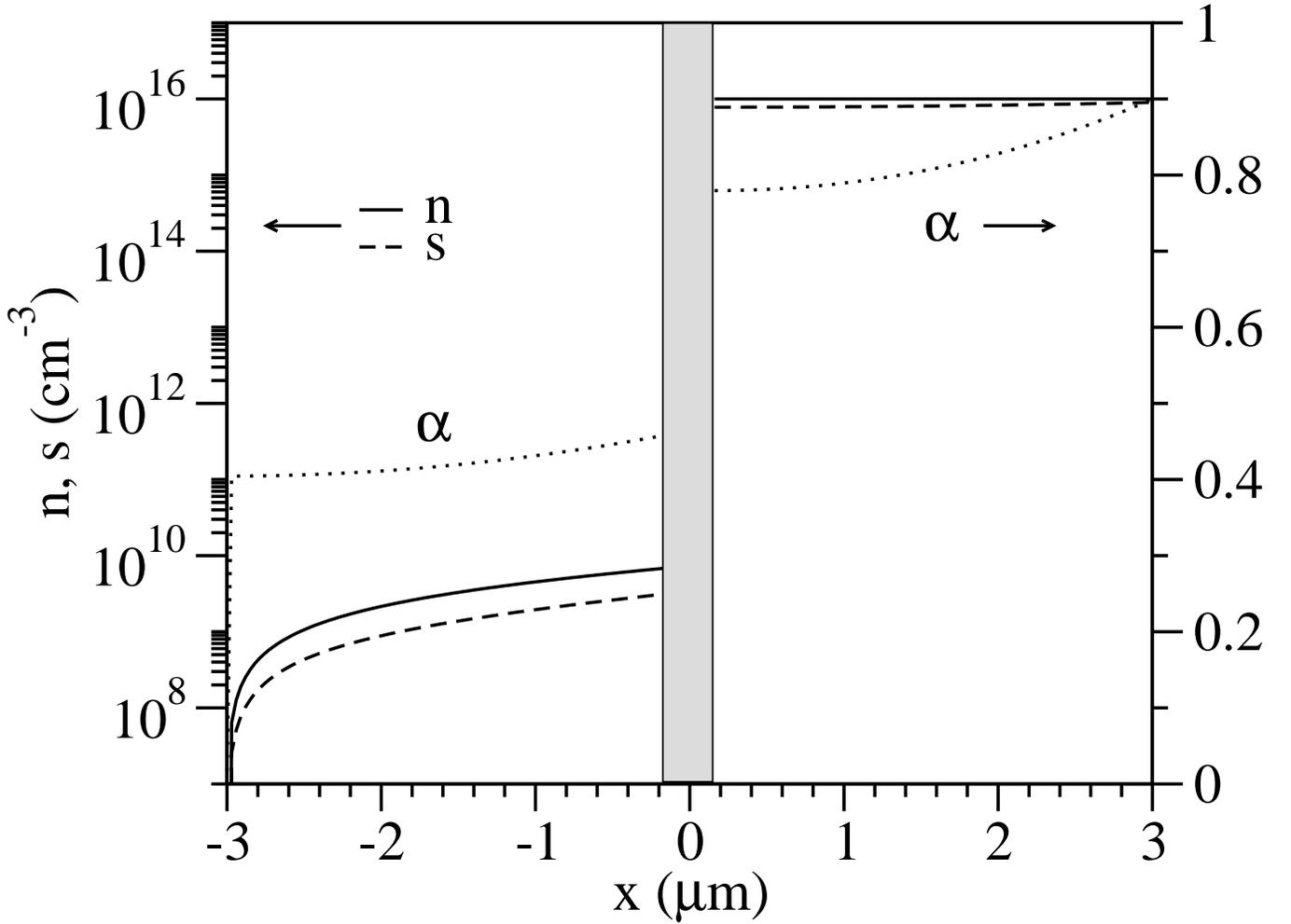}}
\caption{An example of the majority carrier spin injection through the
depletion layer (shaded).  Shown are the spatial profiles of the electron (solid) and
spin (dashed) densities in the magnetic {\it p-n} junction described in the text,
with $\alpha_{0R}=0.5$, $\delta\alpha_{n}=0.4$ (the $p$ region is
nonmagnetic and the left electrode remains Ohmic), and forward bias V=+0.8 volt
($\approx 32$ k$_B$T). The left vertical axis is for the densities, while the right axis is for the 
spin polarization, which is represented by the dotted lines labeled with $\alpha$.
}
\label{fig:2}
\end{figure}

Under the low injection conditions nonequilibrium spin cannot build up
in magnetic {\it p-n} junctions, as was shown in Secs.~\ref{sec:shockley} 
and \ref{sec:continuity}. 
Only if a nonequilibrium (source)
spin is externally injected into either region of the junction,
spin injection through the depletion layer is possible. Here 
we consider the case with a magnetic $n$ side ($\alpha_{0R}\ne 0$)
and a nonmagnetic $p$ side ($\alpha_{0L}=0$), and inject the source
spin at the right contact (but not by the contact itself),
so that $\delta \alpha_n\ne 0$. The left contact remains Ohmic with 
equilibrium carriers and spin ($\delta n_{p}=\delta s_{p}=0$). The nonequilibrium
spin at the depletion layer in the $n$ region is obtained from
Eq.~(\ref{eq:dsR}) (see also Table \ref{tab:2}) as
\begin{equation} \label{eq:dsR1}
\delta s_R=\delta \alpha_{R}N_d=\frac{\delta s_n}{\cosh(\tilde {w}_n/L_{sn})}.
\end{equation}
This boundary condition for spin at the depletion layer can be physically
formulated by requiring that the spin current of the majority carriers  
vanishes at the depletion layer.\cite{zutic02}
This is quite natural to assume, since the spin current in the $n$ side
is proportional to $\alpha_n N_d$, while the spin current in the
$p$ side is proportional to the much smaller $\alpha_L n_L$. Since 
$J_{sR}=J_{sL}$, we can neglect $J_{sR}$ relative to $J_s$ in the 
rest  of the $n$ region. Eq.~(\ref{eq:dsR1}) then follows.

On the left side of the depletion layer the Shockley 
condition, according to Eqs.~(\ref{eq:nL}) and (\ref{eq:sL}), gives for the electron and
spin densities
\begin{eqnarray}
n_L&=&n_{0L}e^V\left (1-\frac{\alpha_{0R}\delta\alpha_R}{1-\alpha_{0R}^2}\right ),\\
s_L&=&n_{0L}e^V\frac{\delta \alpha_R}{1-\alpha_{0R}^2},
\end{eqnarray}
and for the spin polarization
\begin{equation}\label{aL}
\alpha_L=\frac{\delta \alpha_R}{1-\alpha_{0R}^2-\alpha_{0R}\delta \alpha_R}.
\end{equation}
If the source spin has the same direction of polarization as the
equilibrium spin in the $n$ region, 
the electron density $n_L$, and thus the current through the
junction, is reduced. If they are antiparallel, $n_L$ and the current
are enhanced. Neither spin polarization, $\delta \alpha_R$ nor $\alpha_L$, 
depends on $V$ (except for a small
dependence through $\tilde{w}_n$), 
being the same for forward and reverse biases. 
In nonmagnetic junctions ($\alpha_{0R}=0$), all the nonequilibrium
spin polarization is transferred to the minority region, $\alpha_{L}=
\delta \alpha_{R}$, where the nonequilibrium spin has no effect on charge
and current, since $\delta n_L=n_{0L}\exp(V)$.  
This case has been studied numerically for a realistic
model of a spin-polarized nonmagnetic {\it p-n} junction.\cite{zutic01a}

\begin{figure}
\centerline{\psfig{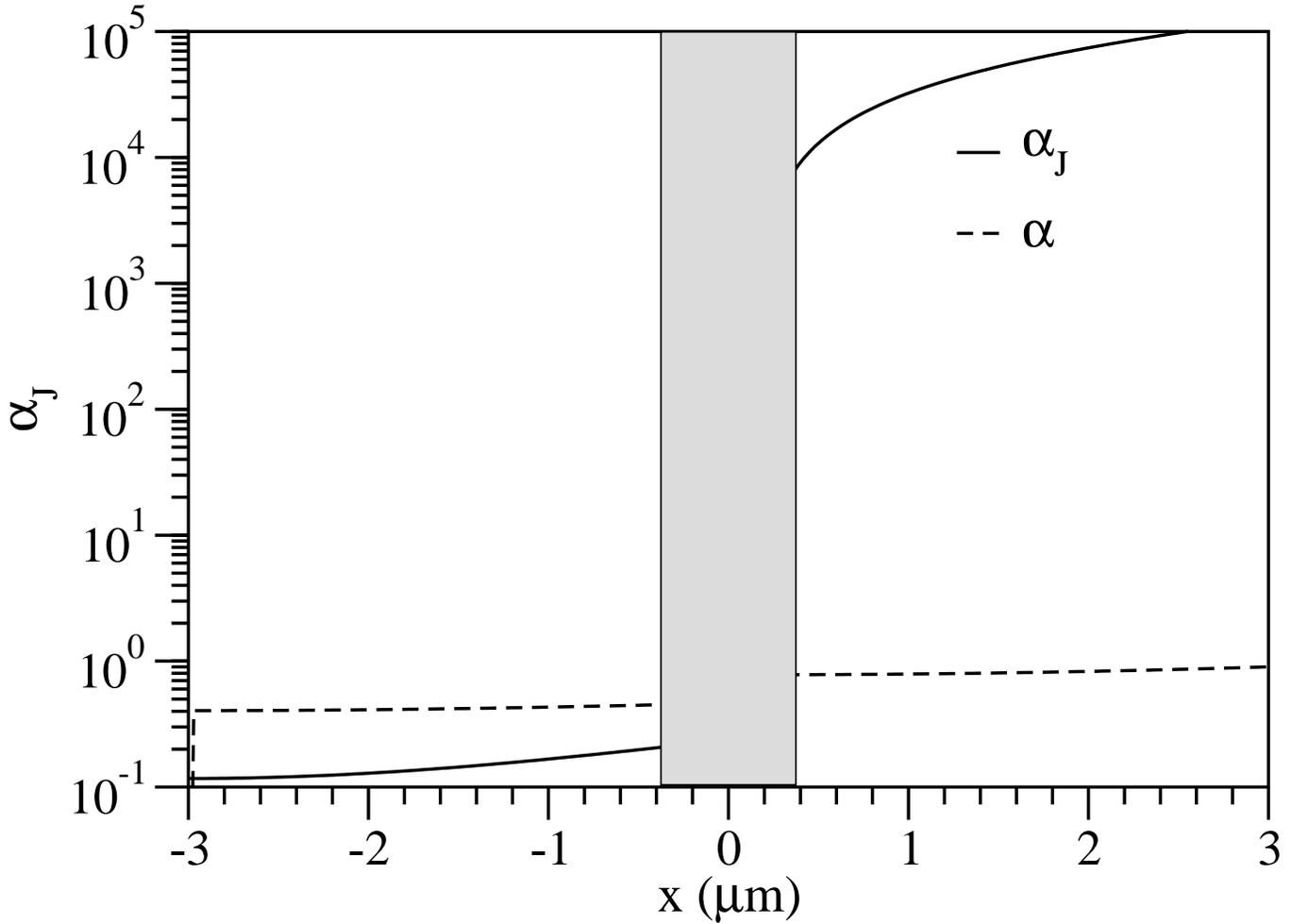}}
\caption{Calculated current spin polarization for the majority carriers spin
injection.
The same parameters as in Fig.~\ref{fig:2} apply.
Both the current spin polarization $\alpha_J$ and the
density spin polarization $\alpha$ are shown for comparison. The current
spin polarization is enormous in the $n$ region, decreasing upon
reaching the depletion layer, and staying smaller than $\alpha$ in
the $p$ region.
}
\label{fig:3}
\end{figure}

The reason for the absence of spin injection through the depletion
layer from a magnetic $n$ region to the nonmagnetic $p$ region,
without a source spin, is the balance between the carrier densities
and thermally activated processes of forward conduction. Let the 
$n$ region be positively magnetized, so that there are more spin up
than spin down electrons. For a nondegenerate statistics, the number
of spin up (down) electrons depends on the spin splitting ($2\zeta_{nn}$) of the
band as $\exp(\zeta_{nn})$ [$\exp(-\zeta_{nn})$]. In the forward transport, electrons
need to be thermally activated to cross the barrier of the built-in
voltage lowered by the external bias. The barrier height is, however,
different for spin up and down electrons. Indeed, spin up (down) electrons
have the barrier higher (lower) by $\zeta_{nn}$, leading to the modulation
of the transport rate by $\exp(-\zeta_{nn})$ [$\exp(\zeta_{nn})$]. These exponential factors
exactly balance the modulation of the carrier densities. As a result, there
is no difference between the transfer rates (density times  
the thermal activation
probability) for the spin up and spin down carriers, the spin up and spin down
currents are equal, and there is no spin current at $R$ (and, by
the continuity of the spin current also at $L$) and thus no spin injection 
into the minority region. 

\begin{figure}
\centerline{\psfig{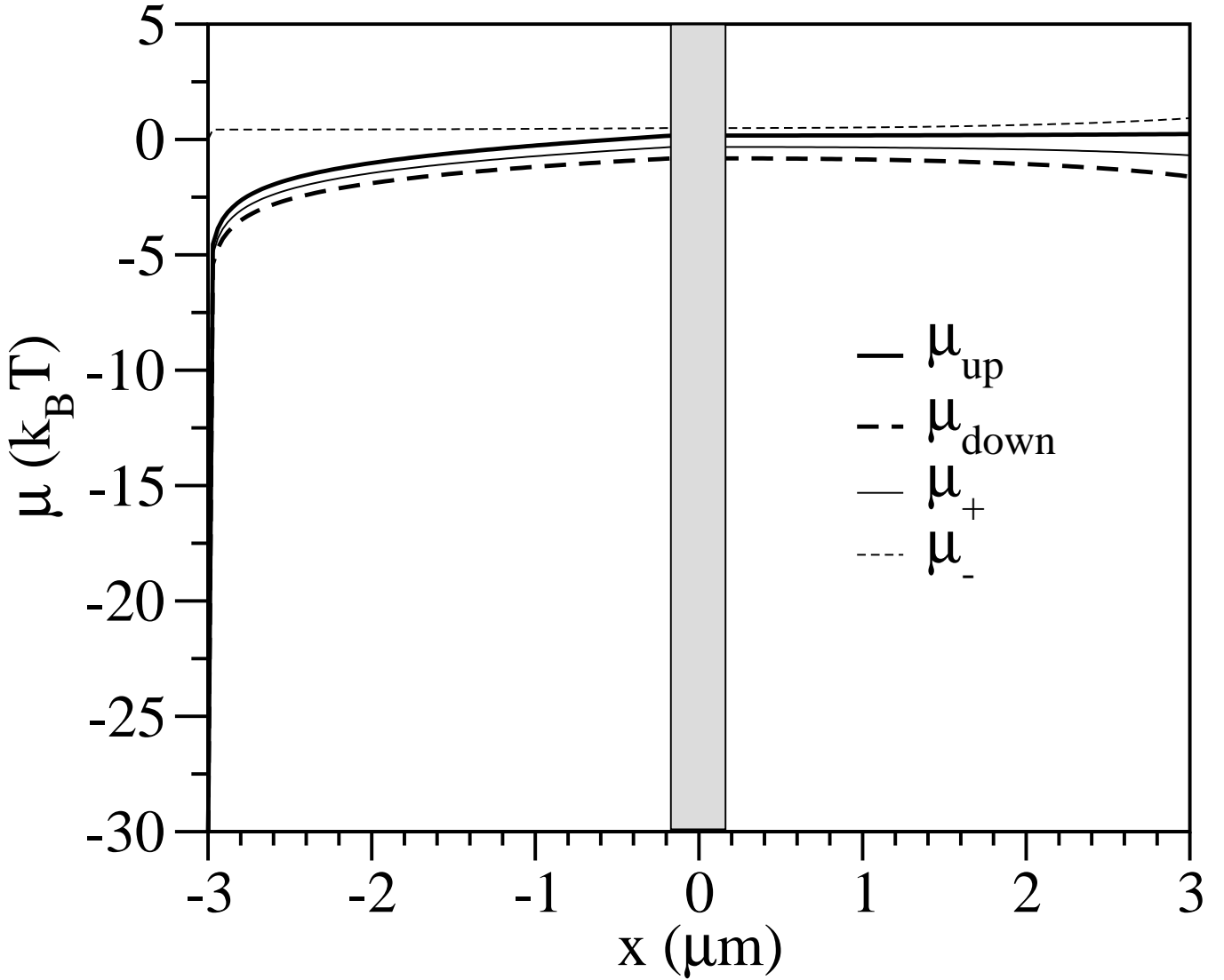}}
\caption{Calculated chemical potential profiles in a magnetic {\it p-n}
junction under the majority carrier spin injection regime. The
same parameters as in Fig.~\ref{fig:2} apply. The chemical potentials
are expressed in the units of $k_B T$.
}
\label{fig:4}
\end{figure}

Figure \ref{fig:2} shows the electron and spin densities, using our
model equations (Table \ref{tab:2}), for the GaAs
magnetic junction example, with $\alpha_{0R}=0.5$ and $\delta \alpha_n=0.4$,
and a forward bias of  $+0.8$ volts.
Spin injection into the minority region is very effective; $\alpha_L$ is 
slightly greater than $\delta
\alpha_R$ [due to the denominator in Eq.~(\ref{eq:aL})]. 
A comparison between the current spin polarization (the profile is
the same as for the spin current $J_s$) and the density spin 
polarization is in Fig.~\ref{fig:3}. The current polarization is
huge at the point of spin injection, since in order to reproduce the
spin polarization $\alpha_n$ by electrical spin injection (which would
depend essentially on $\alpha_J$, see Sec.~\ref{sec:contact}), $\alpha_J$ 
would need to be that large. This is of course not possible, since electrical
spin injection from a ferromagnetic electrode provides $\alpha_J < 1$, since
$\alpha_J$ in ferromagnets is close to the density polarization there.  
The current polarization decreases upon approaching the depletion layer,
since there the spin current decreases in order to be equal to the
spin current at $L$, which is driven by the much smaller density 
of the minority electrons. Figure \ref{fig:4} shows the chemical potential 
profiles for the case. The chemical potentials are chosen to be zero (similarly to $\phi$) at 
$x=w_n$ in the spin unpolarized (but biased) junction, so that at $x=w_p$
they are $-V$ if the contact is Ohmic, as is the present case. This is
the cause of the rapid decrease of the $\mu$'s 
to $\mu=-V$ at $x=-w_p$.
Spin injection in this graph is visible from the finite value of $\mu_-$
(which becomes zero only in the very proximity of the left contact) in the
$p$ region.

\subsection{Spin pumping by the minority carriers} 

\begin{figure}
\centerline{\psfig{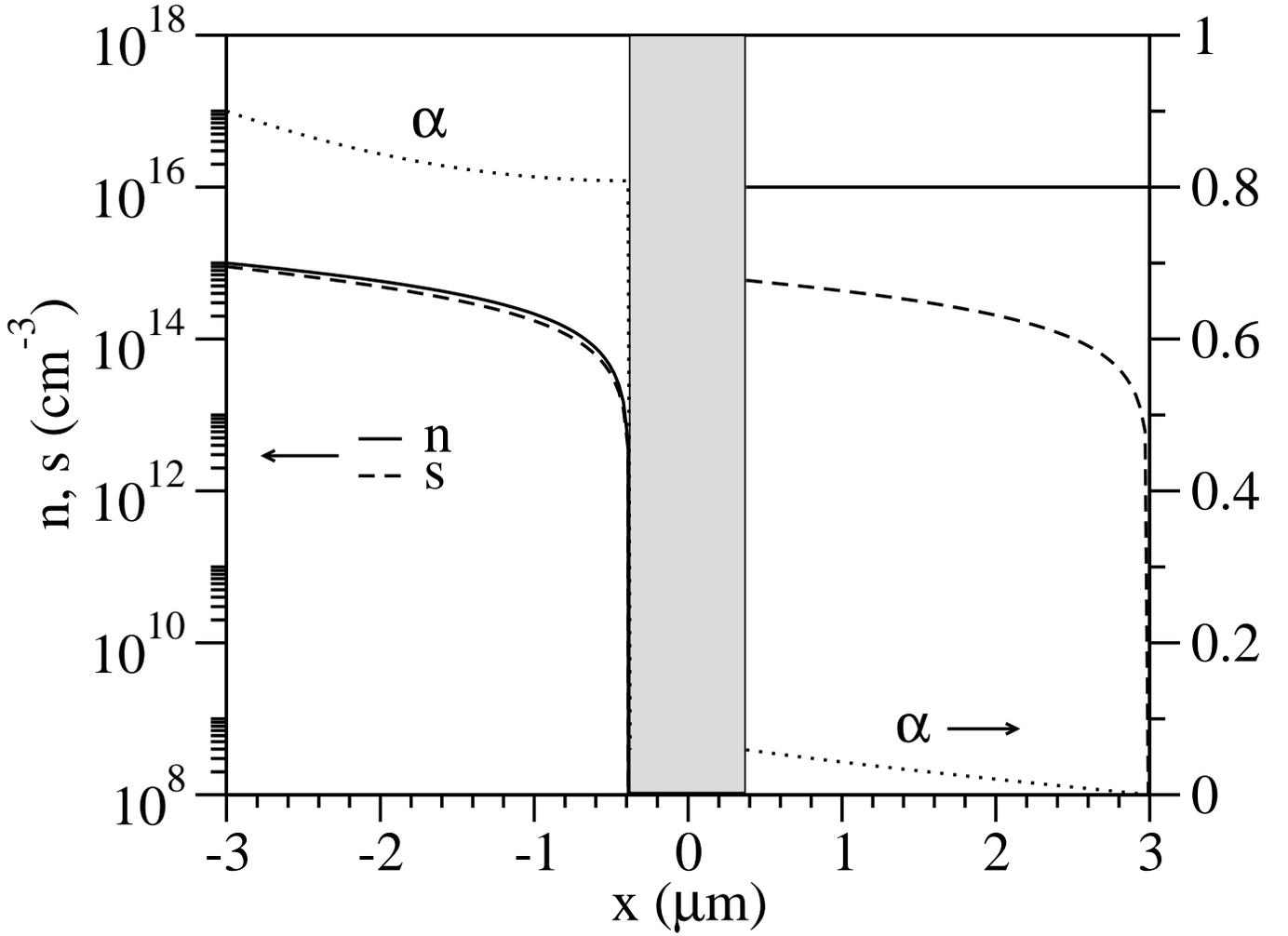}}
\caption{An example of a minority carrier spin pumping through the
depletion layer. The junction is nonmagnetic, but spin-polarized,
and the carrier and spin source is placed at the left electrode,
giving $\delta n_p=1\times 10^{15}$ cm$^{-3}$ and $\delta s_p=1\times 0.9\times 
10^{15}$ cm$^{-3}$ ($\delta\alpha_{p}=0.9$). 
A reverse bias of $-0.8$ volts is applied
(increasing the width of the depletion layer compared to Fig.~\ref{fig:2}).
}
\label{fig:5}
\end{figure}

If large (source) spin density is externally injected
along with the carrier density into the minority region, the
nonequilibrium spin can
reach the depletion layer and be swept by the built-in electric
field to the majority side, where it accumulates.
We have named this effect minority electron spin pumping,~\cite{zutic01a,zutic01b}
since the spin accumulation (which is also a spin amplification,
considering that the resulting spin in the majority region is
much larger than that in  the minority region) depends on the
intensity of the spin current of the minority carriers. The
faster the carriers arrive at the depletion layer, the more 
spin accumulates in the $n$ side. In effect, this is an analogue of the 
optical spin pumping in the majority region,\cite{optical84} except that the
role of circularly polarized light is played by the spin
polarized minority carriers. 

As an illustration consider a nonmagnetic spin-polarized {\it p-n} junction ($\alpha_{0L}=
\alpha_{0R}=0$). Let the carrier and spin densities  
at the left electrode only be out of equilibrium:   
$\delta n_p$, $\delta s_p\ne 0$. 
This happens, for example, when a junction is illuminated by 
circularly-polarized light (like in a spin-polarized
solar cell\cite{zutic01b}) or if the junction is  part of
a spin-polarized junction transistor, in which case the left
electrode simulates the action of the emitter.  
Equation (\ref{eq:dsR}) gives
the ``pumped'' spin polarization in the majority side as
\begin{equation}
\delta s_R=\left ( \frac{D_{np}}{D_{nn}} \right ) \left ( \frac{L_{sn}}{L_{sp}}\right )
\frac{\tanh(\tilde{w}_n/L_{sn})}{\sinh(\tilde{w}_p/L_{sp})} \delta s_p.
\end{equation}
For a large majority region, $\tilde {w}_n \gg L_{sn}$, the injected spin is
(below only holds if $w_p << L_{sp}$)
$\delta s_R \approx (D_{np}/D_{nn})(L_{sn}/\tilde {w}_p)\delta s_p$, 
while for a short majority region,
$\tilde {w}_n \ll L_{sn}$, the injected spin is
$\delta s_R \approx  (D_{np}/D_{nn})(\tilde {w}_n/\tilde {w}_p)\delta s_p$. 
The amount of the pumped spin polarization, 
relative to the amount of the source polarization is 
\begin{equation}
\frac{\delta \alpha_R}{\alpha_p} \approx \left ( \frac{D_{np}}{D_{nn}} \right ) 
\left ( \frac{L_{sn}}{L_{sp}} \right ) \frac{\tanh(\tilde{w}_n/L_{sn})}
{\sinh(\tilde{w}_p/L_{sp})}
\frac{\delta n_p}{N_d}.
\end{equation}
Spin pumping is most effective when the $p$ region is short, $\tilde{w}_p
\ll L_{sp}$, when
\begin{equation}\label{eq:dsR2}
\frac{\delta \alpha_R}{\alpha_p} \approx \left ( \frac{D_{np}}{D_{nn}} \right )
\frac{\min(L_{sn},\tilde{w}_n)}{\tilde{w}_p} \frac{\delta n_p}{N_d}.
\end{equation}
If both $L_{sn}$ and $\tilde{w}_n$ are significantly greater than
$\tilde{w}_p$, the pumped spin (and even the spin polarization)
can be comparable to the source spin (source spin polarization).

\begin{figure} 
\centerline{\psfig{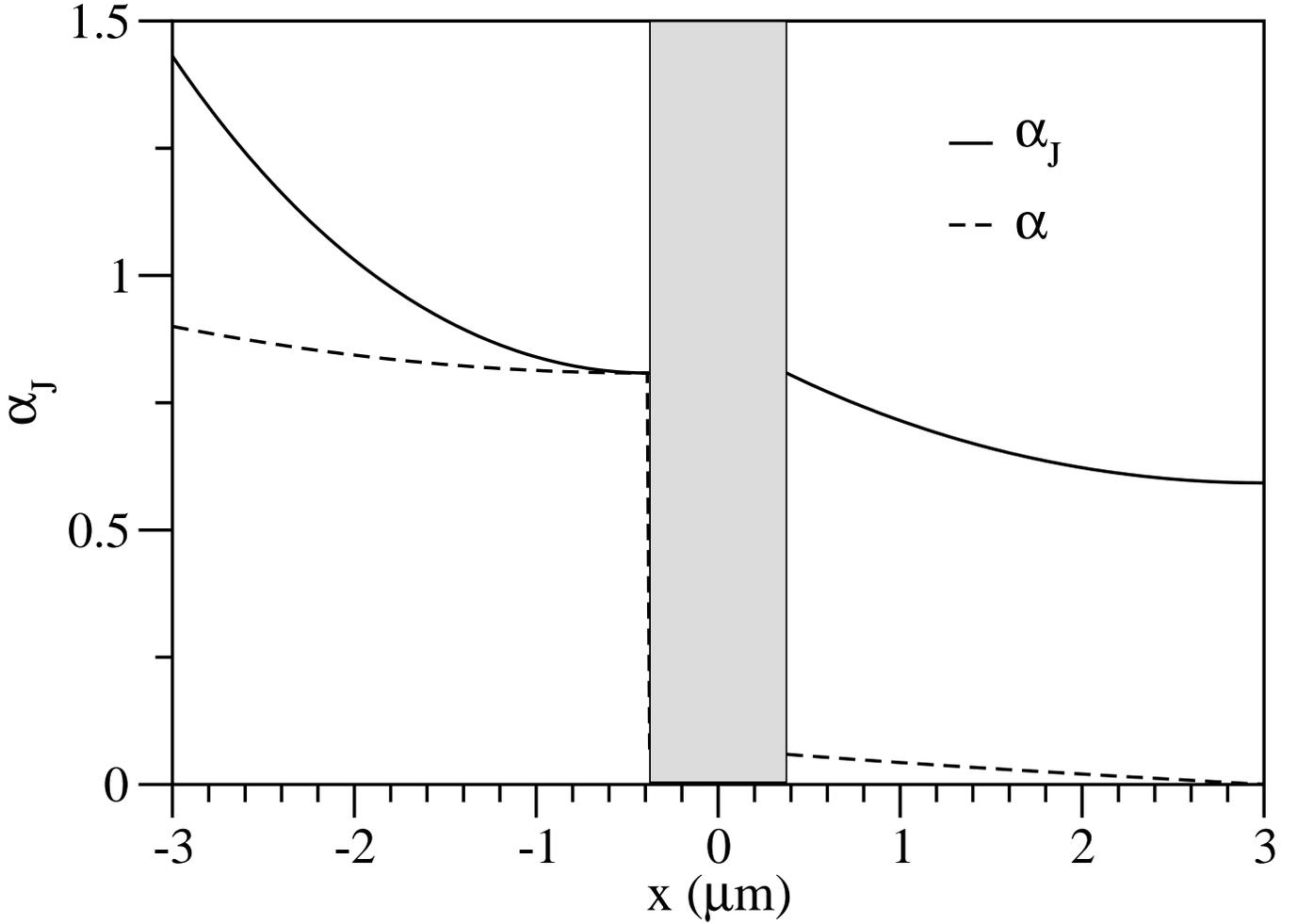}}
\caption{Calculated current spin polarization in the minority spin pumping
regime. Both current, $\alpha_J$, and density, $\alpha$, spin
polarization profiles are shown. The current spin polarization
starts at a value larger than 1 at $x=-w_p$, remains constant 
across the depletion layer where the spin current continuity
is assumed, and decays somewhat in the $n$ region, where its
magnitude is much larger than that of density spin polarization.
}
\label{fig:6} 
\end{figure}

A qualitative argument for the spin pumping is as follows. In the
minority ($p$) side, the spin current goes roughly as
$D_{np} \delta s_p/\tilde{w}_{p}$, where we chose the largest spin in the
region (being the source spin $\delta s_p$) and the smallest 
length scale for the spin decay (here $\tilde{w}_p$). On the $n$ side the 
spin current, along similar reasoning, would be 
$\approx D_{nn} \delta s_R/L_{sn}$, where $\delta s_R$ is the largest spin
in the region and we chose $L_{sn}$ to be the smallest
length scale. Equating the two currents gives Eq.~(\ref{eq:dsR2}). 
Put in words, spin carried by the minority carriers arriving at
the depletion layer is swept into the majority region by the 
large built-in field. In the majority region the spin both diffuses
away and relaxes. In a steady state, the incoming spin flux must equal the outcoming
diffusion and relaxation, which are proportional to the spin density,
so that the greater is the spin influx, the greater the spin 
density.  

\begin{figure}
\centerline{\psfig{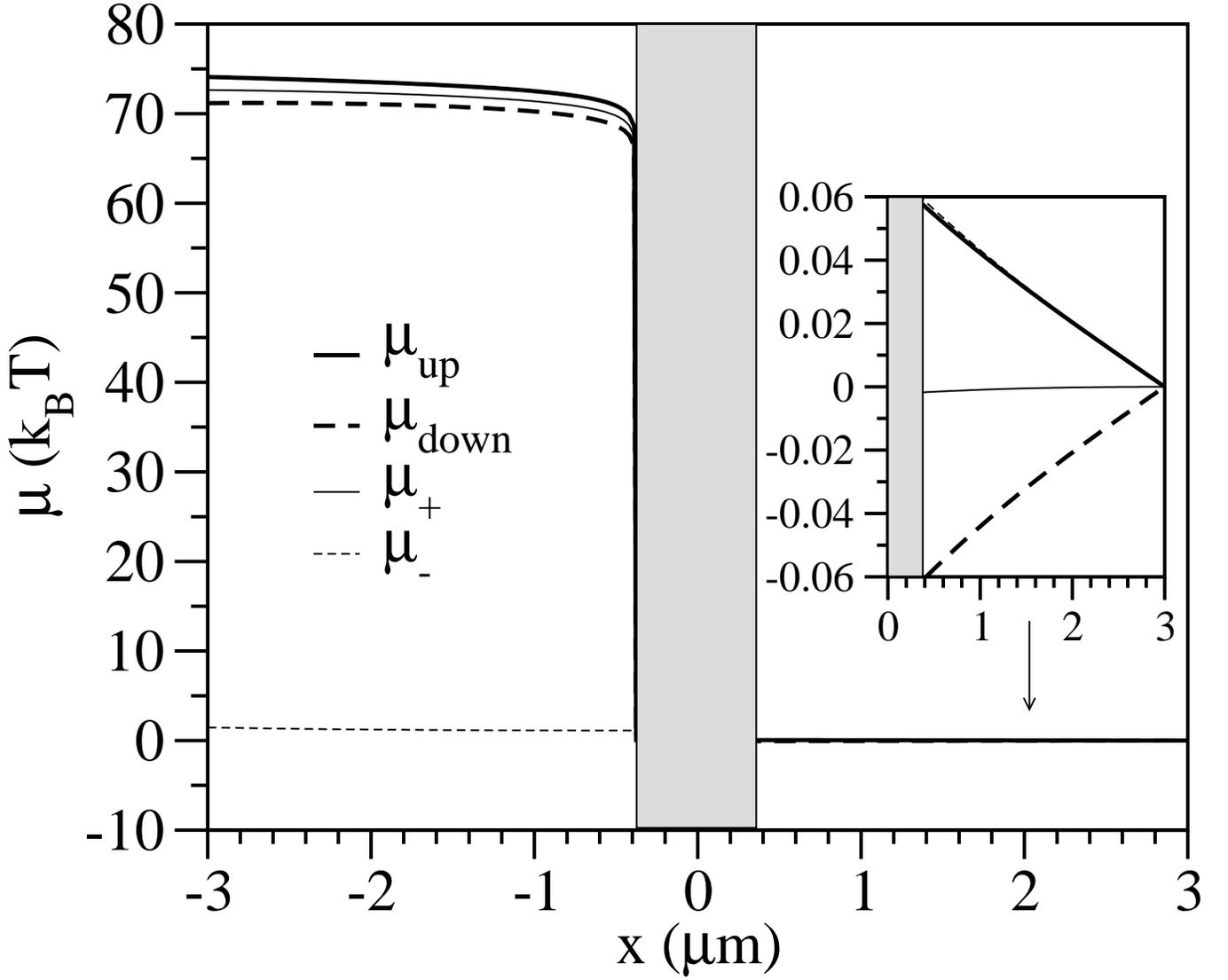}}
\caption{Calculated chemical potential profiles in a nonmagnetic
spin-polarized {\it p-n} junction under the minority carrier spin pumping regime.
The parameters as in Fig.~\ref{fig:5} apply. The input shows the majority
region values on a scale where different $\mu$'s are visible.
}
\label{fig:7}
\end{figure}

A numerical example is shown in  Fig.~\ref{fig:5}. The source carrier and spin
densities are $\delta n_p=10^{15}$ cm$^{-3}$  
and $\delta s_p=0.9\times 10^{15}$ cm$^{-3}$
(the spin polarization $\alpha_p=0.9$). The junction is under reverse bias 
of $-0.8$ volts (note the increase width of the depletion layer compared
to Fig.~\ref{fig:2}). The pumped spin polarization $\alpha_R$ is about 5\%. 
In our numerical example all the length scales involved are comparable (roughly 3 $\mu$m), 
diffusivities uniform ($D_{np}=D_{nn}$) so $\delta s_R\approx \delta s_p$. 
In Fig.~\ref{fig:6} we plot the current spin polarization $\alpha_J$ to 
demonstrate that it significantly differs from the density spin 
polarization $\alpha$. In this example $\alpha_J$ is larger than 1 at
the left electrode due to the chosen boundary conditions, and in the
$n$ region it is much greater than the density spin polarization.
The chemical potential profiles for the case are shown in Fig.~\ref{fig:7}.
In the majority region $\mu_+$ nearly vanishes, while $\mu_{\uparrow} \approx
\mu_{\downarrow}$, demonstrating a positive net nonequilibrium spin 
polarization in the $n$ region. The small magnitudes of the nonequilibrium
chemical potentials in the majority region still yield large spin density,
since they appear in the exponent which multiplies the equilibrium carrier
density, which is large in the majority region (and small in the minority,
where the chemical potentials have accordingly large magnitudes).

\subsection{\label{sec:spinvoltaic}The spin-voltaic effect}

A spin-voltaic effect is a generation of charge emf or 
current by nonequilibrium spin. A first realization of
the spin-voltaic effect was the Silsbee-Johnson spin-charge
coupling\cite{silsbee80,johnson85} in a ferromagnetic/nonmagnetic metal interface
with nonequilibrium spin injected into the nonmagnetic 
metal. The emf across the interface arises due to the
difference in the chemical potentials in the two metals,
with different effects on the different spin states. Analogous phenomena
can occur in many other hybrid systems (semiconductor/metal
or semiconductor/semiconductor). Here we describe a specific
realization of the spin-voltaic effect in magnetic 
{\it p-n} junctions, where the role of the interface is
played by the depletion layer.

Consider a magnetic/nonmagnetic {\it p-n} junction, with 
the $p$ region magnetic ($\delta n_p=0$) 
and the $n$ region nonmagetic but
spin polarized ($\delta \alpha_n\ne 0$). No external bias
is applied ($V=0$). It follows from Eq.~(\ref{eq:dsR}) 
that 
\begin{equation}
\delta \alpha_R=\frac{\delta \alpha_{n}}{\cosh(\tilde{w}_n/L_{sn})},
\end{equation}
which is the same as Eq.~(\ref{eq:dsR1}) (simply expressing the
fact that the polarization is bias independent).
As a result, there will be nonequilibrium carrier and spin
densities in the minority region [see Eq.~(\ref{eq:nL}) and
(\ref{eq:sL})]:
\begin{eqnarray}
\delta n_L&=&n_{0L} \alpha_{0L}\delta \alpha_R, \\ 
\delta s_L&=&n_{0L}\delta \alpha_R. 
\end{eqnarray}
The nonequilibrium minority carrier density $\delta n_L$ leads 
to the minority diffusion and relaxation, and thus to the
charge current (or voltage in an open circuit). The spin-voltaic
current is [see Eq.~(\ref{eq:j1n})] 
\begin{equation}
j_{n}=j_{gn}\alpha_{0L}\delta \alpha_R.
\end{equation}
The current is of the order of the generation  current,
and changes sign with reversing either the magnetic field (which 
reverses $\alpha_{0L}$) or the orientation of the source spin
$\delta \alpha_n$. Neglecting the variation of $\delta \alpha_R$
with bias (through $\tilde{w}_n$), the open circuit voltage for
the spin-voltaic effect is obtained by requiring that $j$
vanishes: 
\begin{equation}
V_{\rm oc}=-\ln \left ( 1+\frac{j_{gn}}{j_{gn}+j_{gp}}
\alpha_{0L}\delta \alpha_R \right ).
\end{equation}
The voltage, which is typically of the order of $k_B T$,
is negative (reverse biasing) if the polarizations are parallel, and 
positive (forward biasing) if they are antiparallel. The spin-voltaic effect here
is similar to the photovoltaic effect, where the photocarriers
generated within the carrier diffusion length $L_{np}$ of the depletion layer are swept
by the built-in field to the majority side, generating photocurrent.
A spin-voltaic effect arises if nonequilibrium spin is generated 
within the spin diffusion length $L_{sn}$ of the depletion layer,
disturbing the balance between the generation and recombination currents.

\begin{figure}
\centerline{\psfig{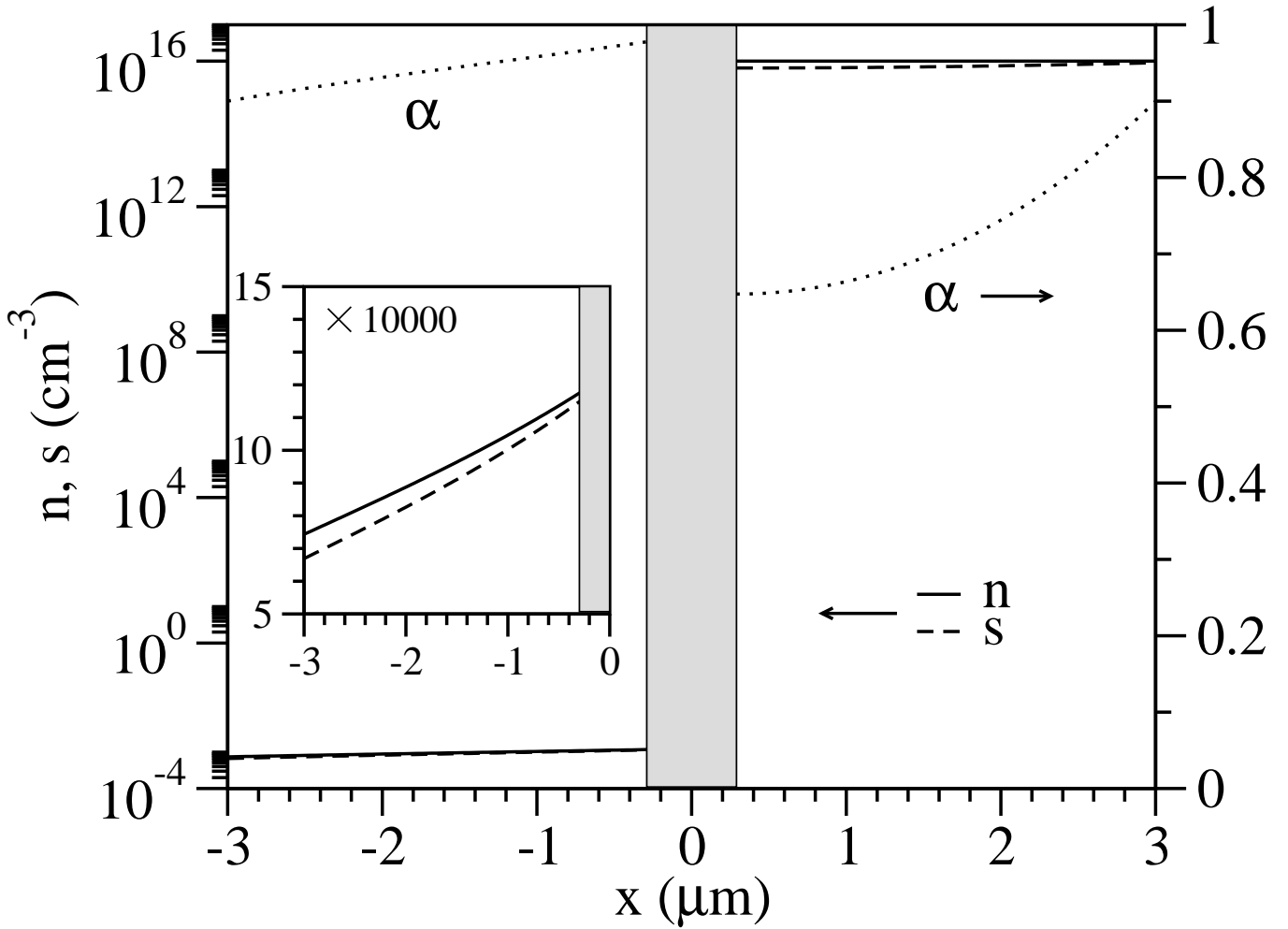}}
\caption{The spin-voltaic effect in a spin-polarized magnetic {\it
p-n} junction. Shown is a junction with a magnetic $p$ region
($\alpha_{0L}\ne 0$) and a nonmagnetic $n$ region ($\alpha_{0R}=0$).
No bias is applied. Both electrodes are Ohmic,
except that there is a spin source at $x=w_n$. In the example
$\alpha_{0L}=+0.9$ and $\delta\alpha_{n}=+0.9$. The carrier
and spin densities in the $p$ region are very close to the
equilibrium values, with a small variation due to the nonequilibrium
spin. The inset shows this variation on a 10000 times increased
scale. Both densities are higher than in equilibrium,
leading to a forward charge current.
}
\label{fig:8}
\end{figure}

\begin{figure}
\centerline{\psfig{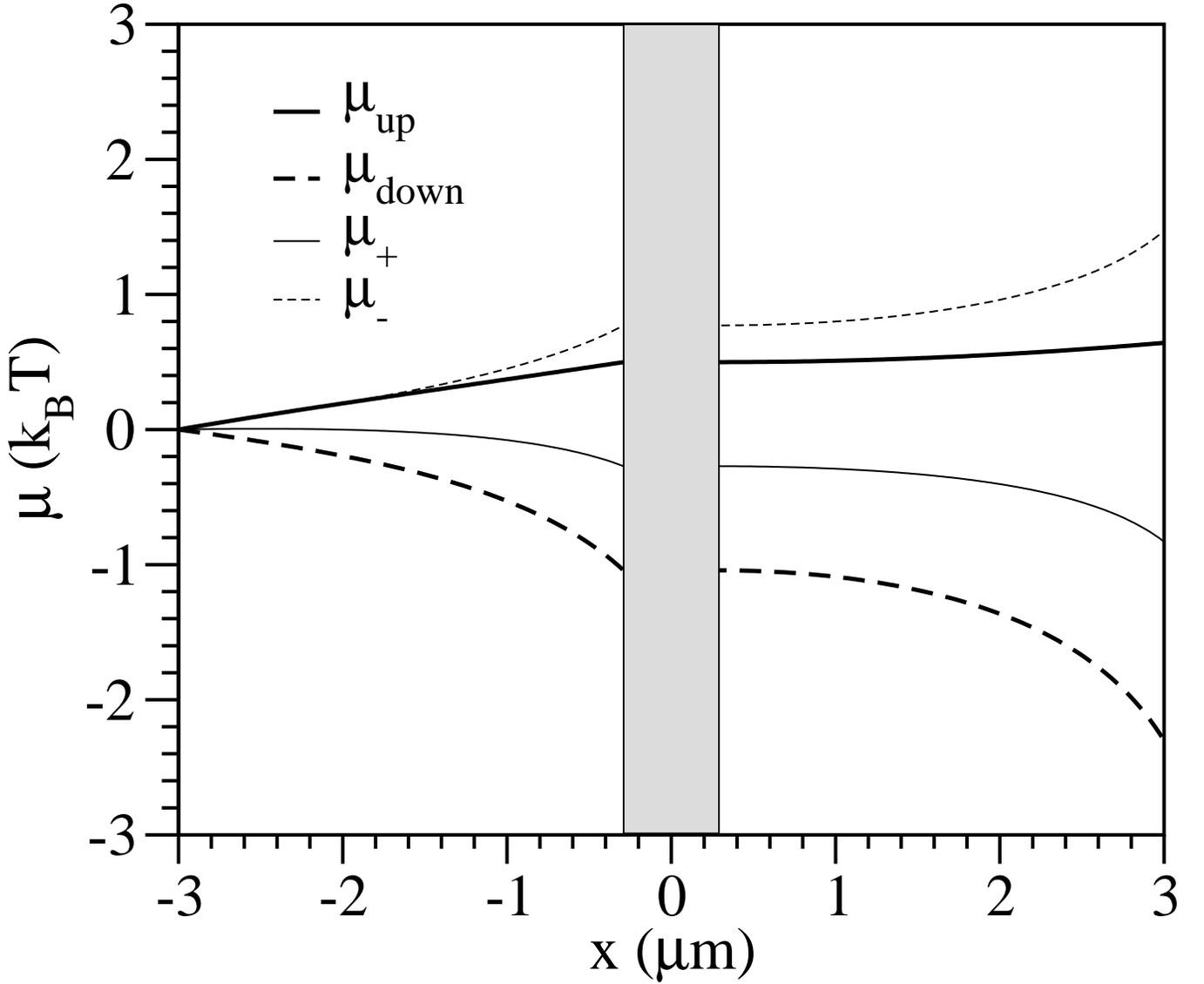}}
\caption{Calculated chemical potential profiles in a spin-polarized
magnetic {\it p-n} junction under the conditions specified in Fig.~\ref{fig:8}.
}
\label{fig:9}
\end{figure}

Indeed, in equilibrium
both the generation and the recombination currents in a magnetic {\it
p-n} junction are equal and there is no net charge flow. 
Let $\alpha_{0L}$ be  
positive. Then the barrier for the majority
electrons to cross the depletion layer (see Fig.~\ref{fig:1}) is smaller for spin 
up than for spin down electrons. If the spin in the majority
region is driven off balance (without applying an external
bias), than the delicate balance of the generation and recombination
currents is disturbed, resulting in a net charge current.
Increasing the number of spin up majority electrons, 
for example, increases the recombination current, since more
electrons have now a smaller barrier to cross (the generation
current does not depend on $\delta \alpha_R$ or bias). In
our geometry, the net electron flow is forward (from the right to
the left, $j_{n}>0$). If, on the other hand, we increase the
number of spin down electrons, more electrons
have now a higher barrier to cross, reducing the recombination
current, resulting in a net reverse flow (from the left 
to the right, $j_{n}<0$). The spin-voltaic effect is the
reason for the giant magnetoresistance of magnetic {\it p-n} junctions,~\cite{zutic02} 
since when a bias $V$
is applied, the spin-voltaic current grows as $\exp(V)$, similarly
to the normal rectification current.

\begin{figure}
\centerline{\psfig{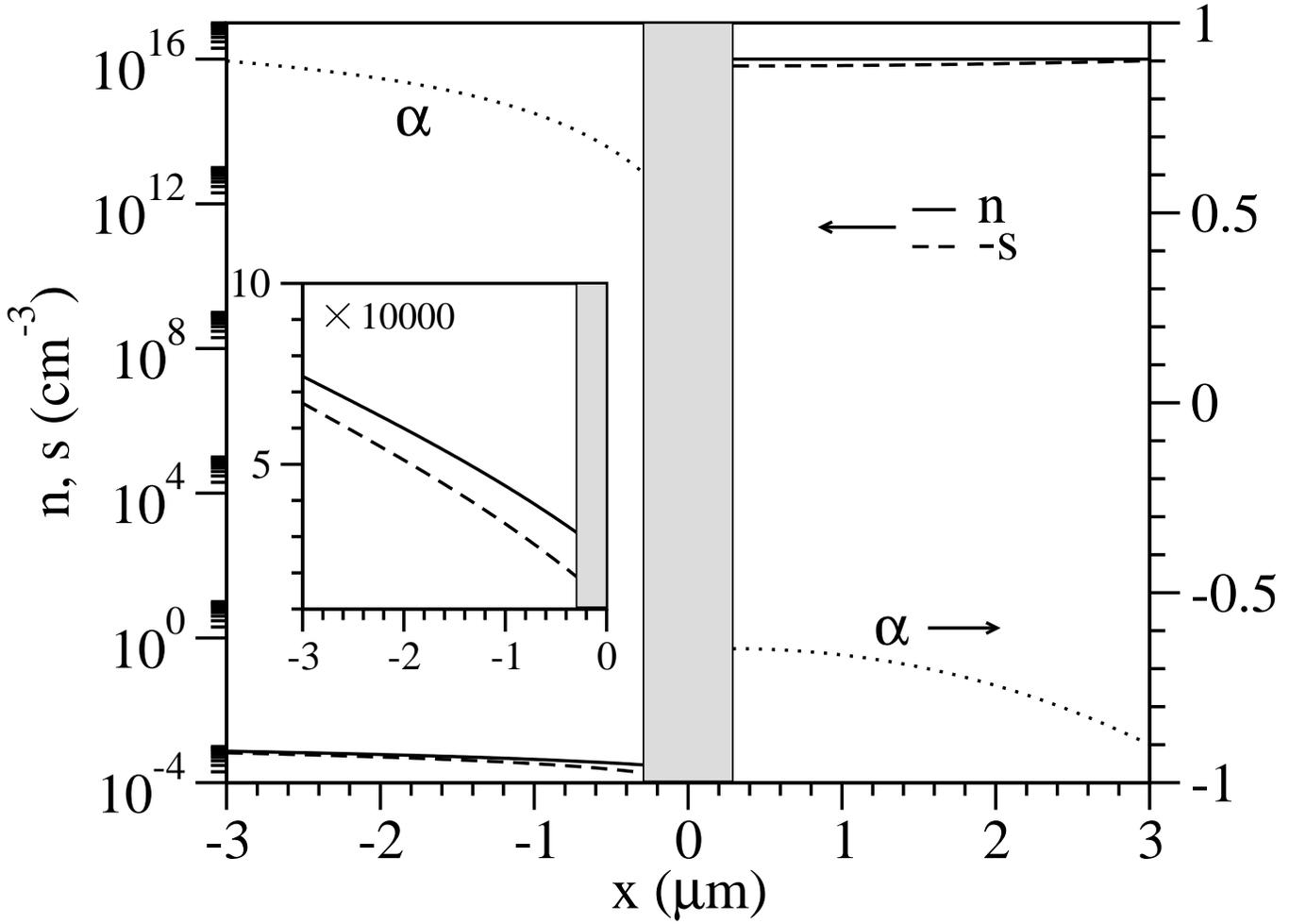}} 
\caption{The spin-voltaic effect in a spin-polarized magnetic
{\it p-n} junction. The same conditions as in Fig.~\ref{fig:8}
apply, but the direction of the source spin is reversed,
$\delta \alpha_n=-0.9$. The figure shows the negative
spin density ($-s$) in the $n$ region (and normal in the
$p$ side). The carrier and spin densities have values close
to the equilibrium ones, but are now somewhat smaller, due
to the presence of the antiparallel nonequilibrium spin.
This density variation, which is seen in the 
inset on a 10000 times greater scale, leads to a reverse
charge flow.
}
\label{fig:10}
\end{figure}

\begin{figure}
\centerline{\psfig{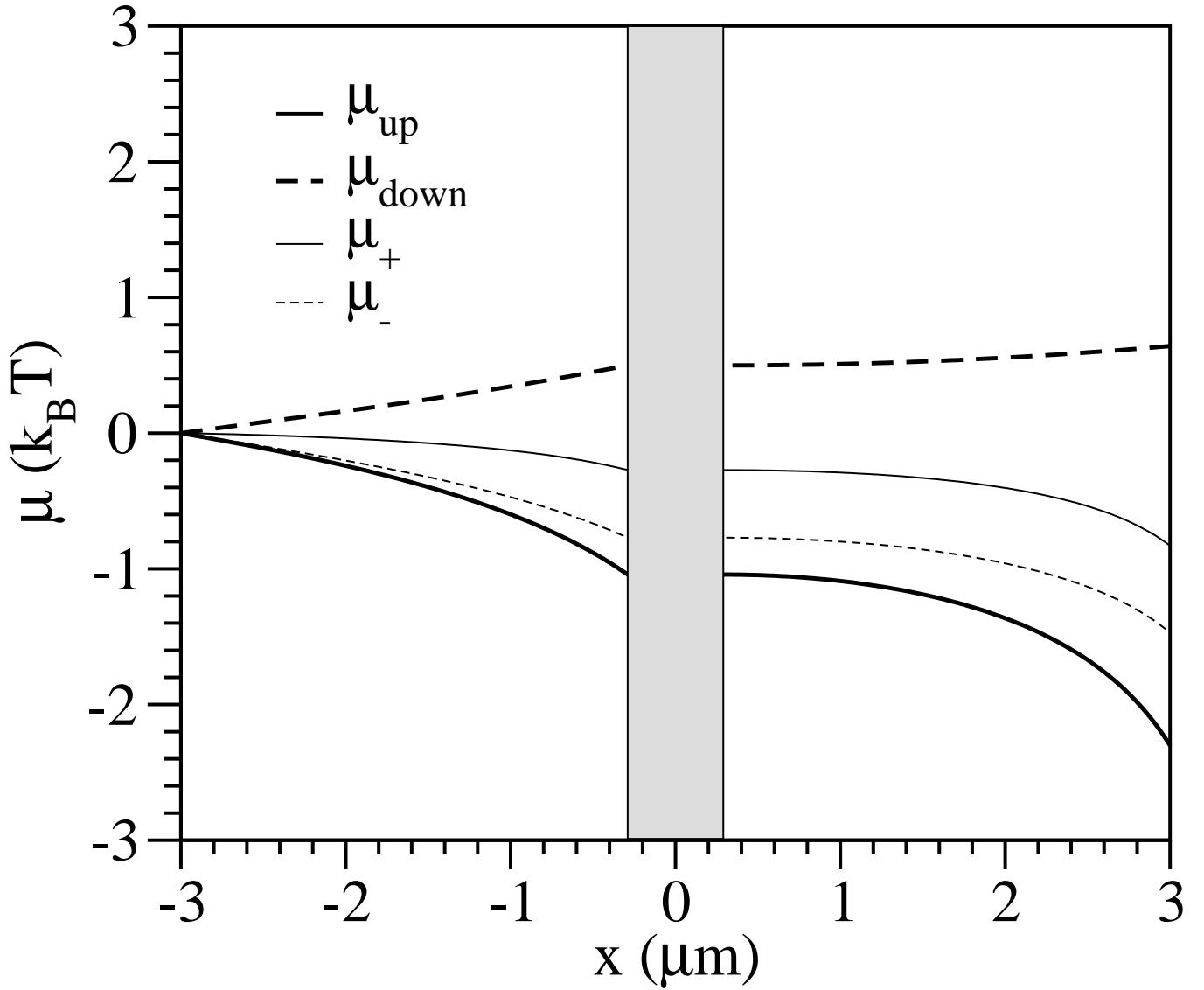}}
\caption{Calculated chemical potential profiles in a spin-polarized
magnetic {\it p-n} junction. The same parameters as in Fig.~\ref{fig:10}
apply.
}
\label{fig:11}
\end{figure}

The spin-voltaic effect is illustrated in Figs.~\ref{fig:8}-\ref{fig:11}. 
First consider parallel spin polarizations, 
$\alpha_{0L}=\delta \alpha_n=+0.9$.
There is no bias, $V=0$. The carrier and spin densities
and the spin polarization are plotted in Fig.~\ref{fig:8}. 
The induced nonequilibrium spin and charge in the
$p$ region are greater than the equilibrium values, 
leading to a forward current of electrons. The spin polarization
is also  higher than in equilibrium. The chemical 
potential profiles are shown in Fig.~\ref{fig:9}. If the spin
polarization of the source spin is reversed, 
$\alpha_{0L}=-\delta \alpha_n=+0.9$, 
the carrier and spin 
densities and the spin polarization decrease in the
minority region, leading to a reverse electron current. 
The density profiles for this case are in Fig.~\ref{fig:10}, and the chemical
potentials are plotted in Fig.~\ref{fig:11}.

Let the magnetic field $B$ controls the conduction band spin splitting.
Then $\alpha_0(B)=-\alpha_0(-B)$. Keeping $\delta \alpha_R$ as an 
independent (of   
$B$) parameter, the direction reversal of the magnetic
field results in a change in charge current:
\begin{equation} 
j_n(B)-j_n(-B)=j_{gn} \frac{\delta n_L(B)-\delta n_L(-B)}{n_{0L}}.
\end{equation}
This is a realization of giant magnetoresistance (GMR) in magnetic
diodes. The relative change of the charge current upon reversing the direction
of magnetic field (while keeping $\delta\alpha_R$ unchanged) can be measured by
the giant magnetoresistance parameter, here denoted as $\beta$:
\begin{equation}
\beta=\frac{\delta n_L(B)-\delta n_L(-B)}{\delta n_L(B)},
\end{equation}
which
at forward bias and $\exp(V) \gg 1$, in terms of the known parameters, can be expressed as
\begin{equation} \label{eq:GMR}
\beta=2\delta \alpha_R \frac{\alpha_{0L}-\alpha_{0R}}{1-\alpha_{0R}^2
+\delta \alpha_R(\alpha_{0L}-\alpha_{0R})}.
\end{equation}
The GMR effect is possible only in magnetically inhomogeneous {\it p-n} junctions
with nonequilibrium spin. As a special case consider the
$p$ region magnetic ($\alpha_{0R}=0$). Then
\begin{equation}\label{eq:beta}
\beta=\frac{2\delta \alpha_R\alpha_{0L}}{1+\delta \alpha_R\alpha_{0L}}.
\end{equation}
This case is a semiconductor analogue of the Silsbee-Johnson
spin-charge coupling,\cite{silsbee80} where a spin emf arises
from the proximity of a nonequilibrium spin in a metal and a
ferromagnetic electrode. A detailed numerical study of the
GMR effect in magnetic diodes can be found in
Ref.~\onlinecite{zutic02}. Putting reasonable parameters
$\delta\alpha_{0L}=0.5=-\delta\alpha_R$, Eq.~(\ref{eq:beta})
gives $\beta=2/3$, which is a $66$ \% GMR.
A more optimistic set, $\alpha_{0L}=0.9=-\delta\alpha_{R}$,
leads to $\beta\approx 8.5$, or a $850$ \% GMR, demonstrating
a great technological potential of magnetic {\it p-n} diodes.

\subsection{\label{sec:contact} Spin injection by the biasing electrode}

Thus far $s_n$ was a free input parameter of
the model. If, however, the biasing electrodes themselves
can inject spin (for example if they are magnetic), then
the source spin density will not be a good starting boundary 
condition. We consider an example of the source spin injection
by the right electrode into the nonmagnetic majority, $n$, region, keeping
only the $p$ region magnetic. We 
assume the model in which the  spin current across the electrode/$n$-region
interface is preserved. In this scenario the boundary condition at $x=w_n$ 
reads (all the current at the contact is carried by electrons, since
the hole density is in equilibrium there)
\begin{equation} \label{eq:eps}
j_{sn}=\alpha_{Jn} j,
\end{equation}
where $j_{sn}=-q J_{s}(w_n)$ 
and $\alpha_{Jn}\equiv \alpha_J(w_n)$ is the spin injection                    
efficiency (here the current spin polarization at the contact) equal, 
in an ideal case, to the spin polarization
in the electrode material reduced by interface
spin relaxation. Our strategy is to convert this
boundary condition to the condition on the spin density: We
calculate $\delta s_n$ as a function of $\alpha_{Jn} j$ and
then use the formulas derived earlier to obtain the
charge current in a self-consistent manner (this is needed
because the boundary spin depends on the current which, in turn, is
calculated using the boundary spin). Equation 
(\ref{eq:Jsn}) gives
\begin{equation}
j_{sn}=\left (\frac{qD_{nn}}{L_{sn}} \right )
\frac{\delta s_n\cosh(\tilde{w}_n/L_{sn})-\delta s_R}{\sinh(\tilde{w}_n/L_{sn})}.
\end{equation}
If we further assume that the left contact is ohmic, by
substituting Eq.~(\ref{eq:dsR1}) for $\delta s_R$, the above equation can be
solved for the source spin density with the result 
\begin{equation}
\delta s_n=\frac{\alpha_{Jn}jL_{sn}}{qD_{nn}}\coth\left(\frac{\tilde w_{n}}{L_{sn}}\right)-
\frac{\gamma_2 s_{0L}\left (e^V-1  \right )}{\tanh^2\left (\tilde{w}_n/L_{sn}\right )}.
\end{equation}
The nonequilibrium spin polarization at the depletion layer then is
\begin{equation}\label{eq:dsRb}
\delta s_R= \alpha_{Jn} j \frac{L_{sn}}{qD_{nn}}\frac{1}{\sinh(\tilde w_{n}/L_{sn})}-
\gamma_4 s_{0L} \left (e^V-1  \right ),
\end{equation}
where another geometric/transport parameter is introduced:
\begin{equation}
\gamma_4=\frac{\gamma_3}{\tanh^2\left (\tilde{w}_n/L_{sn}\right )}.
\end{equation}
In a first approximation one can put $j\approx j_{0n}+j_p$ for the current
in Eq.~\ref{eq:dsRb}. The injected spin $\delta s_R$ is then of the order 
of the minority electron density times $L_{sn}/L_{np}$. This is generally
larger than the spin extraction factor coming from the term with $\gamma_4$, 
but still  small to lead to a significant modulation of nonequilibrium
spin. 
The charge current is obtained by solving Eq.~(\ref{eq:tc}) for $j$, with $\delta\alpha_R$
from Eq. \ref{eq:dsRb}.
The result is
\begin{equation}\label{eq:electrode} \nonumber
j\approx\left (j_p+j_{0n}\right )\left(1+\alpha_{Jn} \hat{\alpha}_{0L}
\right ) -\gamma_4\alpha_{0L}j_{gn}\left (e^V-1\right ) \frac{s_{0L}e^V}{N_d} ,
\end{equation}
where
\begin{equation}
\hat{\alpha}_{0L}=\alpha_{0L} \frac{j_{gn}e^V}{N_d}\frac{L_{sn}}{qD_{nn}}
\frac{1}{\sinh\left (\tilde{w}_n/L_{sn}\right )}.
\end{equation}

Unlike in the case of independent external spin source, the spin
injected by the biasing electrode is very small, because of the
small current flowing in the junction (the current is carried
by the minority carriers), so that only a small spin current can build 
up the source spin. As a result, the spin injected from the 
contact has a very small effect on the charge properties
of the junction. Charge current, in particular, is only slightly 
modified from the spin-equilibrium value of $j=j_p+j_{0n}$. The
spin-voltaic effect is absent (except for the small effect caused 
by spin extraction), since at zero bias no nonequilibrium spin is
injected. Nevertheless, even if small, the I-V characteristics modification 
should be observable at reasonably large biases, and could
be used to characterize spin properties of the junction. 
Furthermore, the above model of spin injection, based on 
Eq.~(\ref{eq:eps}), is rather simple and we cannot exclude the possibility
of a different behavior (especially more pronounced spin dependent
effects) with realistic interfaces. In fact, our method shows
a way how to characterize spin properties of real (electrode) interfaces
by measuring charge response of the junction.

\subsection{\label{sec:biasing}Spin injection and extraction at large biases}

We have shown numerically in Ref.~\onlinecite{zutic02} that spin can be
injected and extracted through the depletion layer at large
biases, even with no source spin present. Significant spin injection
from the magnetic $n$ region into the nonmagnetic $p$
region occurs at large biases and, similarly, significant spin extraction 
occurs from the nonmagnetic $n$ region into the magnetic $p$ region. 
These intrinsic spin injection phenomena have their origin in the low bias
physics. Indeed, there is a (normally negligible) spin injection in 
the absence of source spins even in our theory. We have already demonstrated
spin extraction in Eq.~(\ref{eq:dsR}). If no source spin is present, 
then 
\begin{equation} \label{eq:ex}
\delta s_R=-\gamma_3s_{0L}\exp(V). 
\end{equation}
The nonequilibrium spin in the
$n$ region will have the sign opposite to that of the equilibrium spin
in the $p$ region. The spin is more extracted the larger the bias is. 
Normally, at small biases, the injected spin polarization $\delta \alpha_R=
-\gamma_3s_{0L}\exp(V)/N_d$ is small (in the postulated low injection regime),
but it shows the trend of spin extraction exponentially increasing with
$V$ towards the large bias regime. The reason for Eq.~(\ref{eq:ex}) 
is the
continuity of the spin current across the depletion layer. Without any spin
source, the spin current $J_{sL}$ will be given by the flow of electrons with
the equilibrium spin polarization $\alpha_{0L}$ [see Eq.~(\ref{eq:JsL})]. The
same spin current must appear in the $n$ region at $R$. For positive $s_{0L}$, 
the spin current in $p$ is negative. In the $n$ region, for the spin current
to be also negative there must be a positive gradient of $\delta s$ and,
since $\delta s_n=0$, the spin at $R$ must be negative: $\delta s_R < 0$. 

On the other hand, our theory thus far does not predict any spin injection
from the magnetic $n$ to the nonmagnetic $p$ region. Indeed, in the absence
of source spin, and with $\alpha_{0L}=0$, Eq.~(\ref{eq:dsR}) gives 
$\delta s_R=0$. To explain the intrinsic spin injection observed at 
large biases (but also at small biases, although on a smaller scale\cite{zutic02}),
we have to introduce electric field $E$ into the picture. In fact, once the
nonequilibrium spin becomes small compared to the equilibrium one in the
majority region, even at small biases electric field cannot be neglected.
We will quantify this condition below. There are two factors which need to 
be considered when introducing charge effects in spin transport in the bulk regions. First, 
we will include the electric drift force into the spin current and, second, 
we will explicitly account for charge neutrality by postulating that $n=N_d+
\delta p$ (instead of what was used thus far, $n=N_d$). These two factors
can be normally neglected at small biases, but here we use them to demonstrate
the trends, namely the spin injection, which will become important at 
large $V$. 

Including the $E$-field and the charge neutrality, the spin diffusion equation
from Eqs.~(\ref{eq:Js}) and (\ref{eq:relax}) becomes
\begin{equation}\label{eq:ediff}
\delta s''+E\delta s'=\frac{\delta s}{L_{sn}^2}-\alpha_{0n}\frac{\delta p}{L_{sn}^2}.
\end{equation}
We have neglected the nonlinear terms $\sim sp$, justifiably if 
$(T_1/\tau_{pn})(\delta p/N_d)\ll 1$, which is quite generally the case at low injection. The above
equation needs to be supplemented with the diffusion equation for holes,
\begin{equation} \label{eq:hdiff}
\delta p''=\frac{\delta p}{L_{pn}^2}.
\end{equation}
In Eq.~(\ref{eq:ediff}) the term with the first derivative comes from the electric
drift, while the term proportional to $\delta p$ appears because of the neutrality condition
$n=N_d+\delta p$. The latter term acts as an intrinsic spin source, similarly to the
term of $\delta n$ in the spin diffusion equation (\ref{eq:sdifp}) for the minority electrons. 
The neutrality condition also guarantees that the electric field is uniform
($E'=0$). Equation (\ref{eq:ediff}) has already been considered and 
solved~\cite{yu02,martin02,zutic} without the intrinsic source term, which becomes 
important in bipolar transport at large biases. For completeness, we present the
full solution to Eq.~(\ref{eq:ediff}), as well as the spin current profile, 
in Appendix \ref{appendix:3}.

The full analysis in Appendix  \ref{appendix:3} shows that at least for 
$L_{pn} \ll L_{sn}$, the contribution from the charge neutrality ($\delta n=
\delta p$), that is, from the hole density effects in spin transport, can
be neglected. In the opposite case, the contribution would lower $\delta s_R$,
as can be seen easily by equating $J_{sR}$ in Eq.~(\ref{eq:zetaJsR}) to zero.
The electric field, on the other hand, increases $\delta s_R$, ultimately 
leading to spin injection at large biases. Indeed, from Eq.~(\ref{eq:zetaJsR})
one obtains for spin injection from the majority magnetic region to the minority
nonmagnetic region in the absence of source spin, but at a finite bias,
\begin{equation}
\delta s_R\approx s_{0R} \left (L_{sE}E  \right ) \tanh\left (\frac{\tilde{w}_n}{L_{sE}} 
\right ).
\end{equation}
To obtain $E$ we can use the carrier current continuity across the depletion 
layer: $J_{nL}=J_{nR}$, where $J_{nL}$ is given by Eq.~(\ref{eq:JnL}) and 
$J_{nR}\approx D_{nn}(-N_dE-\delta p_R')$, with $\delta p$  calculated in 
Appendix \ref{appendix:3}. We get
\begin{equation} \label{eq:E}
E\approx \frac{\delta p_R}{N_d L_{pn}}\coth(\tilde{w}_n/L_{pn})+
\frac{D_{np}}{D_{nn}}\frac{\delta n_L}{N_dL_{np}}\coth(\tilde{w}_p/L_{np}).
\end{equation}
The electric field is positive at forward biases, making $\delta s_R$ and thus 
$\delta s_L$ is of the same polarity as $s_{0R}$.
This explains the large bias spin injection (and the increase of spin 
polarization in the magnetic $n$ region) observed in Ref.~\onlinecite{zutic02}.
The electric field spin injection will also happen at small biases, but, 
because the field is very small ($\delta n_L/N_d, \delta p_R/N_d \ll 1$), the
spin injection is negligible. However, the $E$-field must be considered in cases
where spin injection due to source spin leads to $\delta s_R$ as small as
$\delta n_L$.  

Equation (\ref{eq:E}) yields the criterion for neglecting electric
drift in spin transport in the $n$ region (which is in contrast to the
majority carriers, for which electric drift cannot be neglected). Indeed, one needs to compare
the typical magnitudes of the spin drift ($\approx s_{R} E$) with the spin diffusion
($\delta s/L_{sn}$), to obtain 
\begin{equation} \label{eq:cond}
E(s_{0R}+\delta s_R) \ll \delta s_R/L_{sn}.
\end{equation}
For a nonmagnetic $n$ region ($s_{0R}=0$) this is always the case, 
since $EL_{sn}\ll 1$ because
of low injection (and reasonably assuming that the spin diffusion length is
not much greater than the carrier diffusion lengths). For the magnetic region, the above condition is satisfied
if $\delta s_R \gg (EL_{sn}) s_{0R}$, which roughly means that the nonequilibrium
spin in the $n$ region (appearing through the spin source, for example) should
be greater than the nonequlibrium carrier density times the ratio of the 
spin diffusion length the the carrier diffusion length. This is well satisfied
in the low injection regime, where the nonequlibrium carrier densities are
small enough (even if $L_{sn}$ would be one to three orders of magnitude greater
than the carrier diffusion lengths). However, the condition   
(\ref{eq:cond}) 
places the lower limit on the source spin to lead to pure spin diffusion in the
$n$ region.

Finally, the neglected spin relaxation current $J_{s, \rm relax}$ also contributes
to spin injection, more with increasing bias, since then the spin density in the
depletion layer increases and  with it spin relaxation.
The difficulty in introducing $J_{s, \rm relax}$ is that it depends on both bias and
$\delta s_R$ complicating the self-consistent process of obtaining $\delta s_R$ in terms or
the input parameters and bias. One may expect, though, that spin relaxation processes 
in the depletion layer will decrease the spin injection efficiency (that is, reduce 
$\delta s_L$) while allowing for larger $\delta s_R$ to balance the spin current in 
the minority region. Our numerical calculations, which take into account the effects 
of $J_{s,\rm relax}$, find that its contribution is indeed small 
at low biases.\cite{neglect}

\subsection{\label{sec:drift}Magnetic drift in the neutral regions}

Our model and its conclusions thus far were based on magnetic {\it p-n}
junctions with homogeneous magnetic doping in the neutral regions. The doping,
and thus the band spin splitting and the equilibrium spin polarization, 
changed spatially only in the transition region. As a result, the magnetic drift 
force $\zeta'$ dropped from the calculations and the inhomogeneous magnetic doping 
affected the results only through the equilibrium spin densities. Here we take the 
next step and ask how would the physics of magnetic {\it p-n} junctions be affected 
if, additionally, the neutral regions were inhomogeneously doped with magnetic 
impurities (or, to the same effect, were homogeneously doped but placed in an 
inhomogeneous magnetic field). We will show that magnetic drift modifies 
both the spin injection through the depletion layer, and 
the I-V characteristics of magnetic {\it p-n} junctions.
The effects of $\zeta'$ are qualitatively different in the majority and the 
minority regions, so we will discuss the two regions separately. Most of our
discussion below applies equally to homogeneous (in relation to nonmagnetic
doping) semiconductors with spin split majority and minority bands. 

Consider the majority, $n$, region first. In the presence of an inhomogeneous
spin splitting of the conduction band, the electron current in the region is 
\begin{equation}
J_n=D_{nn}(-nE+s\zeta_{nn}'-n').
\end{equation}
The current must vanish in equilibrium where $n\approx N_d$ and $s\approx\alpha_0 N_d$.
This is only possible if a local electric field, 
\begin{equation} \label{eq:E0}
E_0\approx \alpha_0 \zeta',
\end{equation}
develops. The resulting  
electric drift needs to counter
the magnetic drift. The existence of $E_0$ is also
warranted by the vanishing spin ($J_s$) and hole ($J_p$) currents. In the latter the 
electric field needs to balance the equilibrium hole diffusion $\sim p_0'$
[$p_0$ is now spatially dependent through $\zeta_{nn}$, see Eq.~(\ref{eq:appp0n})]. The field 
$E_0$, similarly to the built-in field in the depletion layer,
is an equilibrium field, not an emf, as it does not lead to a net current.

The origin of the equilibrium electric field in otherwise almost homogeneous charge
situation (the majority carrier density is almost constant) can be qualitatively
explained as follows. Take an $n$-type semiconductor doped inhomogeneously with magnetic
impurities in zero magnetic field. In equilibrium the chemical potential is constant.
Switch on a magnetic field. At first, the chemical potential will  
vary with $x$ through $\zeta_{nn}$ according to Eq.~(\ref{eq:mu_0n}). The sample will
come to equilibrium by rearranging its charge as the electrons will move in the direction
of decreasing $\mu_{0n}$, resulting in a constant chemical
potential, but also in a space charge (see below) and a space electric potential opposing further 
electron motion. Then
\begin{equation}
\phi_0=-\ln\cosh(\zeta_{nn}).
\end{equation}
The equilibrium electric field is $E_0=-\phi_0'$, reproducing Eq.~(\ref{eq:E0})
obtained from transport considerations. Electric potential $\phi_0$ bends both the conduction
and the valence band. As for the conduction band, $\phi_0$ tends to straighten the lower 
spin band (say, the spin up band if $\zeta_{nn}$ is positive) and steepen the upper spin
band. At large magnetic fields the band bending of the lower spin band entirely 
eliminates the spatial variations of the band due to $\zeta_{nn}$, while these variations 
are doubled in the upper spin band. The valence band too is affected. Originally
constant, the band acquires spatial variation $-q\phi_0$ to balance the equilibrium
hole diffusion.  

In turn, the inhomogeneous $E_0$ induces space
charge $\rho_0$, according to Poisson's equation:  $\rho_0= E_0'(\epsilon k_B T/q)$. 
In principle, both $E_0$ and $\rho_0$ need to be obtained self-consistently by solving for the
equilibrium semiconductor densities taking into account Poisson's equation
(this was done numerically in Ref.~\onlinecite{zutic02} for the
transition region, where the $E_0$-like field is present due to the inhomogeneous
magnetic doping). However, the induced local charge density is small enough to 
be neglected for most practical purposes (unlike the induced charge density 
in the depletion region). Indeed, the induced changes in the carrier density come to  
$\rho_0/q\approx N_d (\zeta_{nn}'\lambda_{D})^2/\cosh^2(\zeta_{nn})$, where $\lambda_D=
\sqrt(\epsilon k_B T/N_dq^2)$ is the Debye screening length in the majority region.
For GaAs with $\epsilon=13\epsilon_0$ and at room temperature, the density is $\rho_0/q\le 
2\times 10^5 (\zeta_{nn}'[{\rm cm^{-1}}])^2$ cm$^{-3}$. If the magnetic splitting changes by $k_BT$ over 
a micron (so that $\zeta_{nn}'\approx 10^4$/cm), we get $\rho_0\approx 2\times 10^{13}$ cm$^{-3}$. 
This shows that for carrier densities greater than, say, $10^{15}$ cm$^{-3}$ the induced
densities in the carrier concentrations can be neglected, and Eq.~(\ref{eq:E0}) is
a reliable estimate of $E_0$. In general, the space charge can be neglected if the
band splitting varies by $k_B T$ over the length scales greater than $\lambda_D$.  This is in 
complete analogy with space charge considerations due to the usual carrier doping.\cite{tiwari92}
Once $\zeta_{nn}'\lambda_D \agt 1$, which is normally the case when
a magnetic and a nonmagnetic semiconductor form a contact for spin injection, the 
space charge and its diffusion ($\sim \rho_0'/q$) cannot be neglected. Indeed, for 
$\zeta$ changing over a 0.1 $\mu$m, the induced charge density is $q \times 10^{15}$ cm$^{-3}$. 
Selective doping of semiconductors with magnetic impurities on spatial scales both
smaller and 
larger than $\lambda_D$ can prove a useful tool for band structure and 
space charge engineering in designing new spintronic devices.  

Expanding about the equilibrium values for the densities and the electric field, the
electron and spin currents in the $n$ region become
\begin{eqnarray} \label{eq:Jnzeta}
J_n&\approx& -D_{nn}\left (N_d\delta E - N_d\delta \alpha\zeta_{nn}' +\delta p' \right ),\\
\label{eq:Jszeta}
J_s&\approx& -D_{nn}N_d\left (\alpha_0\delta E + \alpha_0\delta \alpha \zeta_{nn}' 
+\delta \alpha'\right ),
\end{eqnarray}
where $\delta E$ describes only the electric field induced by the applied bias. 
Nonequilibrium charge neutrality, $\delta n=\delta p$, is assumed. In a homogeneous
$n$-type semiconductor with finite $\zeta_{nn}'$, maintaining a nonequilibrium spin polarization
would lead to a spin emf according to Eq.~(\ref{eq:Jnzeta}). 
For a constant $\delta \alpha$, for example, the spin emf is 
$-\delta \alpha \Delta \zeta_{nn}$, where $\Delta \zeta_{nn}$ is the drop of $\zeta_{nn}$ 
across the sample. 
Spin injection is modified by the presence of $\zeta_{nn}'$ in Eq.~(\ref{eq:Jszeta}). 
Considering here only a special case of a constant $\zeta_{nn}'$ and large spin polarization, 
$|\alpha_0| \approx 1$, the spin current at $x=d_n$ is modified from Eq.~(\ref{eq:JsR}) to 
\begin{equation}
J_{sR}=-\frac{D_{nn}}{L_{sn\zeta}}\left(\frac{1}{2}\delta s_R \alpha_{0R} \zeta_{nn}' L_{sn\zeta}+F_{\alpha} 
 \right),      
\end{equation}
where the new length scale for spin drift-diffusion is $1/L_{sn\zeta}=\sqrt( \zeta_{nn}'^2/4+1/L_{sn}^2)$
(the spin polarization decay is then governed by two length scales, $L_{sn\pm}$, given 
by $1/L_{sn\pm}=1/L_{sn\zeta}\pm \zeta'/2$),
and spin flux
\begin{equation}
F_{\alpha}=\delta \alpha_n \frac{\exp(\alpha_0\zeta_{nn}'\tilde{w}_n/2)}{\sinh(\tilde{w}_n/L_{sn\zeta})}-
\delta \alpha_R \coth(\tilde{w}_n/L_{sn\zeta}).
\end{equation}
To obtain $\delta s_R$, one can still use Eq.~(\ref{eq:dsR}), but with $L_{sn}$ changed to
$L_{sn\zeta}$ and $\coth(\tilde{w}_n/L_{sn})$ changed to $\coth(\tilde{w}_n/L_{sn\zeta})-
\alpha_{0R}(\zeta_{nn}'L_{sn\zeta})/2$. Similarly for the spin injection at large biases.
Note that in the presence of magnetic impurities $L_{sn}$ will be greatly reduced, so that
$L_{sn\zeta}\approx 2/\zeta'$.

Although the inhomogeneous magnetic doping affects directly only the majority electrons,
it modifies, through $E_0$, transport of the minority holes as well, and thus the I-V characteristics
of the junction. The hole current becomes
\begin{equation}
J_p\approx D_{pn}\left (\alpha_{0R}\zeta_{nn}'\delta p-\delta p' \right ),
\end{equation} 
where the first term describes drift of the nonequilibrium hole density by $E_0$. Together
with the continuity equation for hole current describing electron-hole recombination, the
above equation, again in the limit of a constant $|\alpha_0| \approx 1$ leads to the hole
current at $x=d_n$:
\begin{equation}
J_{pR}\approx D_{pn}\frac{\delta p_R}{L_{pn\zeta}}\left [\frac{\alpha_{0R}\zeta_{nn}'L_{pn\zeta}}{2} +
\coth\left ( \frac{\tilde{w}_n}{L_{pn\zeta}} \right ) \right],
\end{equation}
where we introduced an effective magnetic drift length $1/L_{pn\zeta}=
\sqrt (\zeta'^2/4+1/L_{pn}^2 )$ (the two length scales for the hole density decay are
$L_{pn\pm}$ given by $1/L_{pn\pm}=1/L_{pn\zeta}\pm\zeta'/2$). Since $j_p=qJ_{pR}$, magnetic drift directly 
affects the I-V characteristics of the junction by modifying the hole minority
current. It is in the combination with external bias [$\delta p_R\sim \exp(V)$] that the magnetic drift 
generates current. This effect could be used in electronic detection of magnetic field gradients.

Now we turn to the minority, $p$, region.  
Since the minority electron density can easily accommodate to spatial changes in 
$\zeta_{np}$, no equilibrium electric field is needed  to balance the magnetic
drift force. The carrier and spin currents vanish at $E=0$ for the equilibrium
electron and spin densities, unlike in the $n$ region considered above. 
From Eqs.~(\ref{eq:Jn})-(\ref{eq:relax}), the drift-diffusion 
equations for the minority electrons and spin in the $p$ region are obtained as
\begin{eqnarray}\label{eq:driftn}
n''+En'-\zeta_{np}' s'-\zeta_{np}'' s&=&\frac{\delta n}{L_{np}^2}, \\
\label{eq:drifts}
s''+Es'-\zeta_{np}' n'-\zeta_{np}'' n&=&\frac{\delta s}{L_{np}^2}+
\frac{s-\tilde{s}}{L_{1p}^2}. 
\end{eqnarray}
Transport of minority carriers is thus coupled with the transport of spin. As 
a result, the electron current (and thus the I-V characteristics) will depend
explicitly on nonequilibrium spin and, similarly, spin current will depend
explicitly on nonequilibrium charge. Below we solve Eqs.~(\ref{eq:driftn}) and
(\ref{eq:drifts}) for the specific model of a linear $\zeta_{np}$ (that is, $\zeta_{np}'=
\rm constant$) and in two limits of slow and fast spin relaxation. We will
also neglect the electric field which is by about $\delta n_L/N_d \ll 1$ 
smaller than the inverse of the typical decay length of the densities. 
Magnetic drift brings a new length scale, $L_{np\zeta}$, given by  $1/L_{np\zeta}=
\sqrt(\zeta_{np}'^2/4+1/L_{np}^2)$. The density profiles then decay with 
two length scales, $L_{np\pm}$, which are the inverse of $1/L_{np\zeta}\pm
\zeta_{np}'/2$, depending on whether the diffusion is parallel (minus sign) or 
antiparallel (plus sign) to magnetic drift.

We now consider the limit of vanishing $1/L_{1p}$, which corresponds to slow spin
relaxation (spin diffusion length is the largest length scale in the problem).
We will not present the full density profiles here, only the final results
for the electron and spin currents at the depletion layer boundary $L$, since 
they respectively determine the charge current in the junction and the spin injection 
through the depletion layer.
The boundary conditions and the notation are the same as in Sec.~(\ref{sec:p}). 
The electron current at $L$, in analogy with Eq.~(\ref{eq:JnL}), is
$J_{nL}=-(D_{np}/L_{np\zeta})F_{np\zeta}$, where the modified flux
\begin{eqnarray} \nonumber
F_{np\zeta}&=&\frac{\delta n_L\cosh\left (\tilde{w}_p/L_{np\zeta}\right)-
\delta n_p\cosh\left (\zeta_{np}'\tilde{w}_p/2\right)}
{\sinh\left (\tilde{w}_p/L_{np\zeta}\right )} \\
&-& \delta s_L\frac{1}{2}\left(\zeta_{np}'L_{np\zeta}\right )
-\delta s_p\frac{\sinh\left (\zeta_{np}'\tilde{w}_p/2\right )}
{\sinh \left (\tilde{w}_p/L_{np\zeta}\right)}.
\end{eqnarray}
If the magnetic drift vanishes, $F_{np\zeta}$ becomes $F_{np}$. Since it is
$J_{nL}$ which gives the electron contribution to the total charge current
through the junction, the charge current now explicitly depends on the 
nonequilibrium spin source $\delta s_p$ and the nonequilibrium spin 
at the depletion layer boundary, $\delta s_L$. These contributions will
be important if $\zeta_{np}$ will change on distances smaller than or comparable
to $L_{np}$. Since $J_{nL}$ is sensitive to the sign of $\zeta_{np}'$ (through
the spin contribution), the charge current in a magnetic {\it p-n} 
junction could detect spatial changes in magnetic fields. If the junction
serves as a solar cell or the base of a junction transistor\cite{fabian02}, the nonequilibrium
spin $\delta s_L$ alone will lead to charge current, in analogy with the term
$\delta n_L$ leading to the usual solar cell current. In fact, both the nonequilibrium
spin and carrier densities will be normally present when the junction is illuminated 
by light at $x=-w_p$. The slope of $\zeta_{np}$ then either reduces or enhances
the solar cell current, depending on the sign of $\zeta_{np}'$.

The spin current at $L$ is $J_{sL}=-(D_{np}/L_{np\zeta})F_{sp\zeta}$,
where
\begin{eqnarray} \nonumber
F_{sp\zeta}&=&\frac{\delta s_L\cosh\left (\tilde{w}_p/L_{np\zeta}\right)-
\delta s_p\cosh\left (\zeta_{np}'\tilde{w}_p/2\right)}
{\sinh\left (\tilde{w}_p/L_{np\zeta}\right )} \\
\label{eq:Fspz}
&-& \delta n_L\frac{1}{2}\left(\zeta_{np}'L_{np\zeta}\right )
-\delta n_p\frac{\sinh\left (\zeta_{np}'\tilde{w}_p/2\right )}
{\sinh \left (\tilde{w}_p/L_{np\zeta}\right)}.  
\end{eqnarray}
When neglecting $1/L_{1p}$ in Eq.~(\ref{eq:drifts}) the equations for electrons
and spin become symmetric, so the spin current is obtained from the electron
current by changing $n_p$ to $s_p$ and $n_L$ to $s_L$, and vice versa. Also, 
in our limit of large $L_{1p}$, the effective spin diffusion length is $L_{sn}=
L_{pn}$. In this limit the above equation reproduces $J_{sL}$ from Eq.~(\ref{eq:JsL}). 
If spin relaxation is slow, the spin current in a homogeneously spin split $p$ 
region does not explicitly depend on the electron density. A finite $\zeta_{pn}'$,
however, couples the electron and the spin densities and the spin current acquires 
an explicit dependence on $\delta n_p$ and $\delta n_L$. Spin injection 
is modified by magnetic drift too. If the $n$ region remains magnetically
homogeneous, the injected spin $\delta s_R$ can be obtained by equating
$J_{sL}$ calculated above and $J_{sR}$ from Eq.~(\ref{eq:JsR}). The result
can be written as
\begin{equation}\label{eq:dsRz}
\delta s_R=\gamma_0 \delta s_n-\left (\frac{D_{np}L_{sn}}{D_{nn}L_{np\zeta}}  
 \right ) 
\sinh\left (\tilde{w}_n/L_{sn}\right ) F_{sp\zeta}^0,
\end{equation}
where $F^0_{sp\zeta}$ is  $F_{sp\zeta}$ given by Eq.~(\ref{eq:Fspz}) with
$\delta n_L$ and $\delta s_L$ calculated from Eqs.~(\ref{eq:nL}) and (\ref{eq:sL}) using 
$\delta \alpha_R=0$, that is, $\delta n_L=n_{0L}[\exp(V)-1]$ and 
$\delta s_L=s_{0L}[\exp(V)-1]$. Spin injection is modified in several 
ways. First, there are obvious modifications due to changes in the 
decay lengths, from $L_{np}$ to $L_{np\zeta}$.  
Second, in our limit of $1/L_{1p}=0$ there is no explicit contribution
of $\delta n_p$ to $\delta s_R$ (see Eq.~(\ref{eq:dsR}) with $\gamma_1=\gamma_2$).
Such an explicit dependence appears now because of the magnetic drift. Since
the factor with $\delta n_p$ in Eq.~(\ref{eq:dsRz}) changes sign with $\zeta_{np}'$,
spin injection can be reduced or enhanced. Finally, the large bias spin
extraction will be affected, since it now depends not only on $s_{0L}$ but
also on $n_{0L}$. The latter factor again enhances or reduces the large
bias spin injection depending on the slope of $\zeta_{np}$ (more precisely,
on the sign of $\zeta_{np}'s_{0L}$).

In the opposite limit of fast spin relaxation (which is perhaps  more 
realistic in magnetically doped samples under consideration) one can assume for the
spin to follow the local carrier density changes: $s=\alpha_0 n$. Only 
the drift-diffusion equation for electrons, Eq.~(\ref{eq:driftn}), needs
to be solved in this case. To simplify the discussion, we further assume
that the homogeneous part of the magnetic spin splitting is large, 
and $\alpha_0\approx \pm 1$, with $\alpha_0'\approx 0$. The carrier and spin 
currents have the same magnitude, only the sign can differ if $\alpha_0=-1$.
It thus suffices to look at the carrier current. In analogy with the 
previous case, the spin current is determined by $F_{np\zeta}$, which 
now reads
\begin{eqnarray}\nonumber
F_{np\zeta}&=&\delta n_L \left [\coth \left (\tilde{w}_p/L_{np\zeta}\right )-
\left (\alpha_0 \zeta_{np}' L_{np\zeta}\right )/2   \right ]  \\
&-& \delta n_p\frac{\exp\left (\alpha_0\zeta'\tilde{w}_p/2\right) }
{\sinh\left (\tilde{w}_p/L_{np\zeta}\right )}.  
\end{eqnarray}
The spin current and the spin injection (that is, $\delta s_R$) are then given as 
in the previous limit of slow spin relaxation, but with $F_{sn\zeta}=
\alpha_0F_{np\zeta}$. As in the case of slow spin relaxation, here too the I-V 
curve becomes explicitly dependent on magnetic drift. The strength of the magnetic
drift is determined by the parameter $\zeta_{np}'L_{np\zeta}$, while the sign 
(whether it will enhance or reduce the charge current) on the sign of $\alpha_0\zeta_{np}'$. 
The solar cell current coming from $\delta n_p$ depends exponentially on $\zeta_{np}'$. The
same applies to spin injection. 

\section{\label{sec:summary}summary and conclusions}

We have studied spin-polarized bipolar transport in magnetic {\it p-n} junctions
under the general conditions of applied bias and externally injected (source) 
spin. We have
introduced a model, by generalizing the successful Shockley model of
nonmagnetic {\it p-n} junctions, to include spin-split bands
and nonequilibrium spin. The model is valid only at low injection (small
biases), although it shows the trends of what to expect at large biases
as well. Our theory gives the carrier and spin density profiles in 
the bulk regions (away from the depletion layer),
and explicitly formulates the boundary conditions for
the densities at the depletion layer. In analogy with the original
Shockley model we employ the condition of (quasi) thermal equilibrium across
the depletion layer even when a bias is applied {\it and} a nonequilibrium 
spin is injected. However, the spin polarized case requires an
additional condition to obtain all the relevant input
parameters. This condition we formulate in terms of the continuity
of the spin current across the depletion layer.  The obtained boundary
conditions allow us to generalize the standard diode formulas to the
case of spin-polarized magnetic diodes, resulting in  a new formulation of
the I-V characteristics. Although to explain the physics
of bipolar spin-polarized transport we use spin polarized
electrons only, we also 
give all the formulas needed to calculate
the I-V curves for spin polarized holes as well (in the Appendix, where we also
show how the equilibrium properties of {\it p-n} junctions
are modified in the presence of spin-split bands).  

We have applied our theory to several cases which we believe are
important for spintronics. We demonstrate that only nonequilibrium
spin can be injected across the depletion layer. Effective spin injection
from a magnetic into a nonmagnetic region, without a source
spin, is not possible at small biases. We show how this
claim is relaxed at large biases, which build up a nonequilibrium 
spin in the magnetic majority region, and then inject this spin into the nonmagnetic 
minority region. Similarly, we demonstrate that spin can be extracted
at large forward biases from the nonmagnetic majority region to the
magnetic minority one. We also study spin injection by the minority
carriers to the majority region. Physically, this process can
be described as spin pumping, since the resulting
accumulation/amplification of spin in the majority 
region depends on the spin current of the minority carriers.
The accumulated spin can be greater than the source spin,
which in effect is a spin amplification. 
A realization of the spin-voltaic effect is found at  
the interface (here the depletion layer) between the minority
magnetic region ($p$) and the nonmagnetic but spin-polarized majority
region. The spin-voltaic effect 
is demonstrated by the generation of charge current 
by nonequilibrium spin (at no applied bias). This is
also a spin-valve effect, 
since the direction of the
charge current can be reversed by reversing an applied 
magnetic field.  
The spin induced nonequlibrium charge
density is also the basis for the spin capacitance of the
spin-polarized junctions\cite{zutic01b} as well as
for the spin and magnetic field dependent charge capacitance
of magnetic {\it p-n} junctions.\cite{fabian02}
Next we have studied 
(source) spin injection by the biasing electrode and
shown that this is not a very effective means of spin injection,
at least for a simple model considered.  Finally, we demonstrated
that if the neutral regions have nonequilibrium band spin 
splitting, the resulting magnetic drift can significantly 
affect both the I-V characteristics of the junction and the
junction spin injection capabilities. 

Our theory is general enough to be applicable 
to various semiconductor spintronic devices operating under the
conditions of small injection and nondegenerate carrier statistics.
While we have already demonstrated the extensive generality 
of the theory by applying it to a large number of specific 
model device simulations, we envisage many more potential 
spintronic junction devices where our models will be useful.
Such devices can be, for example, bipolar  spin junction transistors\cite{fabian02} 
or spin thyristors,
with great technological potentials, and where charge 
currents (and their amplification) can be controlled not only by bias, but also by 
nonequilibrium spin and magnetic field. However, to apply the
theory to realistic device structures,  many physical 
aspects of the model will need to be modified. In 
many cases the spin states of the carriers are not simple
spin doublets, but rather multiplets, as a result of
the spin-orbit coupling. In addition, the electron-hole
recombination is, in general, spin selective, so if
both electrons and holes are spin polarized, more realistic
models for the recombination need to be introduced.
Furthermore, carrier recombination and spin relaxation
depend on the carrier density, an effect which may be
found important if ferromagnetic semiconductors are
employed. Other possible additions to the model may include
a realistic treatment of spin relaxation (and carrier
recombination) in the depletion layer and finite spin relaxation
at the contact electrodes. Structural modifications may
include inhomogeneous magnetic doping (or inhomogeneous
magnetic fields) also in the bulk regions, and schemes
based on two or three dimensional spin bipolar transport.
Since, at the moment, there is a lack of experimental understanding
of bipolar spin transport, theoretical modeling (both 
analytical as presented here or numerical, which is of greater
applicability, as reported in 
Refs.~\onlinecite{zutic01a,zutic01b,zutic02}) is particularly 
important. We believe that although quantitative aspects
of spin-polarized bipolar transport may be seriously modified,
our theory captures the essential physics and the predicted
phenomena are robust enough to be present in more realistic
situations.

\acknowledgments{We would like to thank Prof. E. I. Rashba 
for useful discussions. This work was supported by DARPA, the 
US ONR, and the NSF.}

\appendix

\section{\label{appendix:1}
Equilibrium properties of magnetic {\it p-n} junctions}

To study equilibrium properties of magnetic {\it p-n}
junctions we consider both electrons and holes 
spin polarized. Denote the electron and hole spin densities as
$s_n$ and $s_p$, and reserve the second subscript (if needed)
to denote the region. Symbol $0$ denotes the equilibrium values.
As in the main text, the energies (potentials) are given 
in the units of $k_B T$ ($k_B T/q$). 
Further, 
denote as $2\zeta_n$ and $2\zeta_p$ the spin band splittings of
the conduction and valence bands.  
We adopt the convention 
that $\zeta_n$ ($\zeta_p$) 
is positive when the spin up electrons 
(holes) have a lower energy than those in the spin down states.
Both electrons and holes are assumed in thermal equilibrium, 
obeying nondegenerate Boltzmann statistics.

The equilibrium carrier densities obey the law of mass action, 
now reading
\begin{equation}
n_0p_0=n_i^2\cosh(\zeta_n)\cosh(\zeta_p).
\end{equation}
As a result, the minority carrier densities (electrons in the
$p$ region and holes in the $n$ region) are
\begin{eqnarray}
n_{0p}&=&\frac{n_i^2}{N_a}\cosh(\zeta_{np})\cosh(\zeta_{pp}), \\
\label{eq:appp0n}
p_{0n}&=&\frac{n_i^2}{N_d}\cosh(\zeta_{pn})\cosh(\zeta_{nn}).
\end{eqnarray}
Similarly, the corresponding equilibrium spin densities are
\begin{eqnarray}
s_{0,np}&=&\frac{n_i^2}{N_a}\sinh(\zeta_{np})\cosh(\zeta_{pp}), \\
s_{0,pn}&=&\frac{n_i^2}{N_d}\sinh(\zeta_{pn})\cosh(\zeta_{nn}),
\end{eqnarray}
so that  the equilibrium spin polarizations of electrons
and holes are $\alpha_{0n}=\tanh(\zeta_n)$ and
$\alpha_{0p}=\tanh(\zeta_p)$. The built-in voltage, which is the 
electrostatic potential
drop across the depletion layer depends on the band
splittings (thus on the equilibrium spin polarizations):
\begin{equation}
V_b=\ln\left (\frac{N_aN_d}{n_i^2}\right)-
\ln\left[\cosh(\zeta_{pp})\cosh(\zeta_{nn})  \right ].
\end{equation}
The built-in voltage is slightly reduced by the 
spin splitting. Note that only the band splittings of the majority
carriers affect the built-in field. The reason is that the chemical
potentials in the bulk regions (considered separately) depend
only on $\zeta_{nn}$ and $\zeta_{pp}$:
\begin{eqnarray}\label{eq:mu_0n}
\mu_{0n}&=&\mu_{i}+\ln\left(\frac{N_d}{n_i}\right)-\ln\cosh(\zeta_{nn}),\\
\mu_{0p}&=&\mu_{i}-\ln\left(\frac{N_a}{n_i}\right)+\ln\cosh(\zeta_{pp}).
\end{eqnarray}
Here $\mu_i$ is the chemical potential for the intrinsic (undoped)
and unpolarized case. In making a junction, the built-in
field arises upon equilibrating the two chemical potentials: $V_b=
\mu_{0n}-\mu_{0p}$. The band splitting does not affect the nondegeneracy
of the carrier statistics, since the distance between the chemical
potential and the lower conduction (upper valence) spin
band does not change with $\zeta$ at large $\zeta$.

\section{\label{appendix:2}Spin polarized holes}

Spin polarization of holes can be treated separately from that 
of electrons, since, in our model, electron and hole transport
are independent (only minority diffusion is considered), and
the electron-hole recombination is spin independent (in our simplified
picture electrons of one spin can recombine with holes of
either spin). Inclusion of spin polarization of holes into our
theory then amounts to simple notation exchange, $p$ with $n$
and $L$ with $R$. For completeness, we present all the
important formulas which are needed to obtain the charge current
contribution by spin polarized holes. Since this is a separate
section from the main text, we adopt the same notation for
the hole spin as we had before for electrons, without using more
elaborate set of indexes. The hole spin density (only in this section) is $s$ and
the hole spin polarization is $\alpha$. All the other symbols retain
their original meaning.

In analogy with Eq.~(\ref{eq:jn}), the hole charge current is
\begin{equation}\label{eq:jp}
j_p=j_{0p}+j_{1p}+j_{2p},
\end{equation}
where 
\begin{eqnarray} \label{eq:j0p}
j_{0p}&=&j_{gp}\left (e^V-1\right ), \\ \label{eq:j1p}
j_{1p}&=&j_{gp}e^V \delta \alpha_L\frac{\alpha_{0R}-\alpha_{0L}}{1-\alpha_{0L}^2}, \\
\label{eq:j2p}
j_{2p}&=&j_{gp}\frac{1}{\cosh(\tilde{w}_n/L_{pn})}\frac{\delta p_n}{p_{0R}}.
\end{eqnarray}
The hole generation current is
\begin{equation}
j_{gp}=\frac{qD_{pn}}{L_{pn}}p_{0R}\coth\left (\tilde{w}_n/L_{pn}\right ).
\end{equation}
The (now majority) hole spin, in analogy with Eq.~(\ref{eq:dsR}) can be expressed
as
\begin{equation}
\delta s_{L}=\gamma_0\delta s_{p}+\gamma_1 \delta \tilde{s}_{n}
+\gamma_2\alpha_{pn}^0\delta p_n-\gamma_3s_{0R}\left (e^V-1\right ),
\end{equation}
where the geometric/transport factors are
\begin{eqnarray}
\gamma_0&=&1/\cosh(\tilde{w}_p/L_{sp}), \\
\gamma_1&=&\left (\frac{D_{pn}}{D_{pp}} \right )\left
(\frac{L_{sp}}{L_{sn}} \right )
\frac{\tanh(\tilde{w}_p/L_{sp})}{\sinh(\tilde{w}_n/L_{sn})},\\
\gamma_2&=&\left (\frac{D_{pn}}{D_{pp}} \right )\left
(\frac{L_{sp}}{L_{pn}} \right )
\frac{\tanh(\tilde{w}_p/L_{sp})}{\sinh(\tilde{w}_n/L_{pn})},\\  
\gamma_3&=&\gamma_2\cosh(\tilde{w}_n/L_{pn}).
\end{eqnarray}
Finally, the (now minority) hole density and spin in the $n$ side of the depletion layer are
\begin{eqnarray} \label{eq:pR}
p_R&=&p_{0R}e^V\left
( 1+\delta \alpha_L \frac{\alpha_{0R}-\alpha_{0L}}{1-\alpha_{0L}^2}\right ), \\
\label{eq:spR}
s_R&=&s_{0R}e^V\left (1+\frac{\delta \alpha_L}{\alpha_{0R}}\frac{1-\alpha_{0R}\alpha_{0L}}
{1-\alpha_{0L}^2} \right ).
\end{eqnarray}
Physical consequences of the spin polarization of holes in bipolar transport 
are in complete analogy with the physics discussed in the main text 
where only spin-polarized electrons are considered. In particular, the hole charge
current $j_p$ from Eq.~(\ref{eq:jp}) needs to be substituted to the
total charge current formula, Eq.~(\ref{eq:tc}). In many cases, however, one can 
realistically treat only one carrier type as spin polarized. If, for example, holes 
have a very short spin lifetime (or small diffusivity), their spin polarization
(even their contribution {\it per se}) does not need to be considered. The 
exceptional cases are the large bias spin-polarized transport and  magnetic drift 
in the bulk regions, treated in Secs.~\ref{sec:biasing} and \ref{sec:drift}, respectively, 
in which electron and hole transport can be strongly coupled.

\section{\label{appendix:3} Majority electron drift and diffusion}

The spin profile in the $n$ region is affected by the electric field and 
charge neutrality, as described by Eq.~(\ref{eq:ediff}). Assuming the same
boundary conditions for spin as in Sec.~\ref{sec:n}, that is, $\delta s(d_n)=
\delta s_R$ and $\delta s(w_n)=\delta s_n$, and the boundary conditions
for holes $\delta p(d_n)=\delta p_R$ and $\delta p(w_n)=0$ (Ohmic contact), the
solution to Eq.~(\ref{eq:ediff}) can be written in the form analogous to 
Eq.~(\ref{eq:dsn}):
\begin{equation}\label{eq:dsR3}
\delta s = e^{-E(x-d_n)/2}\left(\delta \hat{s}_R \cosh\eta_{sE}+
F_{sE}\sinh\eta_{sE}\right )+\alpha_{0n}\delta \hat{p}.
\end{equation}
We now describe the new notation. The effective deviations from the equilibrium
of spin and hole densities are
\begin{eqnarray}
\delta \hat{s}_R&=&\delta s_R-A\alpha_{0n}\delta p_R\left 
[1+\lambda_{E}\coth\left (\tilde{w}_n/L_{pn}\right )  \right ], \\ 
\delta \hat{p}&=&A\left [\delta p+\lambda_E \left (L_{pn}/D_{pn}\right )J_p  \right ],
\end{eqnarray}
where $\lambda_E=E/\kappa^2L_{pn}$ measures the strength of the electric field
for drifting spin, $\kappa=\sqrt (1/L_{pn}^2-1/L_{sn}^2)$ (the singular case of
$\kappa=0$ is excluded from the solution), and $1/A=-(L_{sn}\kappa)^2
(1-\lambda_E^2)$. The normalized flux is
\begin{equation}
F_{sE}=\frac{-\delta \hat{s}_R\cosh\left (\tilde{w}_n/L_{sE} \right ) +
\delta \hat{s}_n\exp\left (E\tilde{w}_n/2\right ) }{\sinh\left (\tilde{w}_n/L_{sE}\right )},
\end{equation}
introducing a length scale $L_{sE}$ for electric spin drift: $1/L_{sE}=\sqrt (E^2/4 + 1/L_{sn}^2)$,
which is also used to define $\eta_{sE}\equiv (x-d_n)/L_{sE}$. This length scale
was already introduced in Refs.~\onlinecite{yu02,martin02,zutic}.  The new effective spin 
density 
\begin{equation}
\delta \hat{s}_n=\delta s_n-\frac{A\alpha_{0n}\delta p_R \lambda_E}{\sinh\left 
(\tilde{w}_n/L_{pn}\right )}.
\end{equation} 
Finally, the hole density profile
$\delta p$ is obtained by solving independently for hole diffusion, Eq.~(\ref{eq:hdiff}). For 
completeness we show the result:
\begin{equation}
\delta p=\delta p_R \cosh\eta_{pn}+F_{pn}\sinh\eta_{pn},
\end{equation}
where $\eta_{pn}\equiv (x-d_n)/L_{pn}$ and
\begin{equation}
F_{pn}=-\delta p_R \coth\left ( \tilde{w}_n/L_{pn} \right ).
\end{equation}
The hole current is then $J_p=-D_{pn} \delta p'$. 

The importance of electric drift for the majority electron spin transport is in aiding 
the spin injection through the depletion layer, from the majority
magnetic to the minority nonmagnetic region. To see how spin can be injected through
the depletion layer we need to know the spin current at the depletion layer boundary. The
spin current profile is $J_s=-D_{nn}(sE+s')$, where $s$ is given by Eq.~(\ref{eq:dsR3}). 
The spin current at $x=d_n$ is
\begin{eqnarray} \nonumber
J_{sR}&=&-D_{nn}(s_{0n}+\frac{1}{2}\delta s_R)E \\ \nonumber
&+&\frac{D_{nn}}{L_{sE}}
\frac{\delta s_R\cosh \left (\tilde{w}_n/L_{sE}\right )-
\delta s_n\exp\left(E\tilde{w}_n/2\right)}{\sinh(\tilde{w}_n/L_{sE})} \\
\label{eq:zetaJsR}
&+&\frac{D_{nn}}{L_{pn}}\alpha_{0n}\delta p_R A  \coth\left 
(\frac{\tilde{w}_n}{L_{pn}}\right ) \\ \nonumber
&-&\frac{D_{nn}}{L_{sE}}\alpha_{0n}\delta p_R A  
\coth\left (\frac{\tilde{w}_n}{L_{sE}}\right ),
\end{eqnarray}
where we neglected terms of order $\lambda_ED_{nn}\delta p_R/L_{pn}$. For most practical 
cases in magnetic {\it p-n} junctions the electric field at low injection can be neglected,
so that $EL_{pn}, EL_{sn} \ll 1$. Then $L_{sE} \approx L_{sn}$, $\kappa\approx 1/L_{pn}$, 
and $A\approx -(L_{pn}/L_{sn})^2$. Since $E$ is of order $\delta p_R/L_{pn}$, the contribution
to the spin current (and thus to spin injection) from the hole density is negligible, since
in the considered limit $A\ll 1$. If $\delta s_n$ is greater than, say,  $10^{-3} \times N_d$, 
then 
also the contribution from the electric drift can be neglected (not limited to the above
limit), verifying our theory in the main text. If, however, the source spin is small, and there
is appreciable forward bias, the electric drift has to be taken into account for describing 
spin injection across the depletion layer. This is done in Sec.~(\ref{sec:biasing}).




\begin{thebibliography}{}
\bibitem{note} Typically one region would be nonmagnetic (or only weakly magnetic).
This would
be the region where the nonequilibrium spin would be generated and
transported to the depletion layer. If the region were magnetic, such 
transport could be hindered by fast spin relaxation.
\end{thebibliography}
\end{document}